\documentclass[conference, letterpaper]{IEEEtran}

\usepackage{cite}
\usepackage{amsmath,amssymb,amsfonts}
\usepackage{algorithmic}
\usepackage{graphicx}
\usepackage{textcomp}
\usepackage{xcolor}
\usepackage{multirow}
\usepackage{multicol}

\usepackage{epsfig}
\usepackage{times}
\usepackage{float}
\usepackage{afterpage}
\usepackage{amsmath}
\usepackage{amstext,cite}
\usepackage{amssymb,bm}
\usepackage{latexsym}
\usepackage{color}
\usepackage{graphicx}
\usepackage{amsmath}
\usepackage{amsthm}
\usepackage{graphicx}
\usepackage[center]{caption}
\usepackage{pstricks}
\usepackage{caption}
\usepackage{subcaption}
\usepackage{booktabs}
\usepackage{multicol}
\usepackage{lipsum}% dummy text
\usepackage[T1]{fontenc}
\usepackage{hyperref}
\usepackage{aecompl}
\usepackage{mathrsfs}

\usepackage{tikz}
\usetikzlibrary{matrix,calc}
\usepackage[space]{grffile}

\newtheorem{thm}{\bf Theorem}
\newtheorem{rem}{\bf Remark}

\usepackage{mathtools}
\DeclarePairedDelimiter\ceil{\lceil}{\rceil}
\DeclarePairedDelimiter\floor{\lfloor}{\rfloor}

\begin{document}
\title{Optimal Linear Coding Schemes for the Secure Decentralized Pliable Index Coding Problem} % (PICOD) 

\author{%
\IEEEauthorblockN{%
Tang Liu and Daniela Tuninetti, \\
University of Illinois at Chicago, Chicago, IL 60607 USA, \\
Email: {\tt tliu44, danielat@uic.edu}
}%
% \thanks{%
%    The results of this paper were presented in parts at %the following conferences:
%    the 2019 IEEE  International Symposium on Information Theory, Paris, France.
%    The results were also made available online %as of Nov. 2015 as follows:  
%    at arXiv:1904.05272.
%    The work of the authors was supported in part by the National Science Foundation under award number 1527059.%
% }
}
\maketitle

\begin{abstract}
    We study the secure decentralized Pliable Index CODing (PICOD) problem with circular side information sets at the users.
    {The security constraint forbids every user to decode more than one message while decentralized setting means there is no central transmitter in the system.}
    Compared to the secure but centralized version of the problem, a converse bound from one of our previous works %\cite{centralized_secure_picod} 
    showed a factor of three difference in optimal code length under the constraint of linear encoding.
    In this paper, we first list the {\it \textbf{linearly infeasible cases}}, that is, problems where no linear code can simultaneously achieve both correctness/decodability and security. 
    Then, we propose linear coding schemes for all remaining cases and show that their attained code length is to within an additive constant gap from our converse bound.
    %Our schemes can achieve the linear converse bounds in~\cite{centralized_secure_picod} within an additive constant gap for all feasible cases.
    %Showing that the linear converse bounds in~\cite{centralized_secure_picod} are optimal within a constant gap.
\end{abstract}

\section{Introduction}
\label{sec:backgroud}

In this paper we study the Secure Decentralized Pliable Index CODing (SD-PICOD) problem. 
PICOD is a variant of the Index Coding (IC) problem, motivated by broadcast systems where users have message side information sets to help reduce the number of transmissions needed to satisfy the users' demands~\cite{index_coding_with_sideinfo}.

\paragraph*{Index Coding}
The traditional IC setting consists of $m$ messages, one central transmitter, and $n$ users. 
The transmitter and the users are connected by an error-free broadcast channel.
Each user has some messages locally stored as its side information set and has one pre-determined message to decode.
The structure of the side information sets and the desired messages are known to the transmitter and all the users.
The transmitter %thus encode based on the knowledge and 
broadcasts %the codewords
coded symbols to all users.
The users decode based on the received %codewords 
coded symbols and their own side information set.
The goal for IC is to find the smallest code-length / number of transmitted coded symbols such that all users can decode correctly their desired messages. %s successfully.

\paragraph*{Pliable Index Coding}
PICOD is a variant of IC motivated by the scenarios where the desired message at the users is not be pre-determined~\cite{BrahmaFragouli-IT1115-7254174} such as for example streaming services and online advertisement systems.
In PICOD a user is satisfied whenever it can correctly decode at least one message that is not in its side information set. 
Therefore, the transmitter can leverage the freedom of choosing the desired messages for the users so as to reduce the code-length. %required number of transmissions. 
Compared to the IC with the same number of users and the same side information sets, PICOD needs less number of transmissions to satisfy all the users~\cite{SongFragouli-ISIT2016sub-7176784}.

\paragraph*{Decentralized Problems}
%\paragraph*{Decentralized Index Coding}
The decentralized IC problem is motivated by peer-to-peer communication systems where there is no central transmitter and instead 
coded symbols are generated by the users based on their side information %in the system 
and sent through a common time-sharing noiseless broadcast channel. %to all the users.
The goal again is to find the minimal code-length %number of transmissions 
that allows every user to correctly decode its desired message.
Under linear encoding constraint, the minimum code-length %number of transmissions for a 
of the decentralized IC is shown to be no more than twice that %of the minimum number of transmissions for 
of its centralized counterpart~\cite{embedded_ic}.
%\paragraph*{Decentralized Pliable Index Coding}
The pliable version of decentralized IC was studied in~\cite{decentralized_picod}, where information theoretical bounds on the optimal code-length were given for several cases.
For the solved cases, the multiplicative gap between centralized and decentralized PICOD is usually much less than two.

\paragraph*{Secure Centralized Problems}
%\paragraph*{Secure Index Coding}
Security in IC %, where the transmitter needs to generate the codewords such 
means that the users can only decode their desired message while all other messages that are not in their side information set must remain unknown to the users.
The secure IC problem was first proposed in~\cite{private_ic}, where %the key system is
private one-time-pad keys were used to %make
meet the security demand. %satisfied.
{A weaker definition of security which can be achieved without security keys has also been discussed in~\cite{private_ic} and has been extended to PICOD setting~\cite{secure_picod_achievability}.}
% In~\cite{secure_picod_achievability} the secure PICOD problem was studied \dt{but with a weaker definition of security compared to~\cite{private_ic} (which turns out to allow to meet the security constraint without the use of security keys).} %a key system. 
Achievable and converse bounds for the case of circular side information structure and linear encoding were derived in~\cite{private_picod}. %In~\cite{private_picod}, a general linearly optimal achievable scheme and corresponding converse bound have been proposed.

\paragraph*{Secure Decentralized Problems}
In this paper we are interested in the secure and decentralized setting.
%\paragraph*{Decentralized Secure Index Coding}
% \dt{Decentralized Secure Index Coding????}
%
%\paragraph*{Decentralized Secure Pliable Index Coding}
Recently, the secure PICOD problem in~\cite{private_picod} has been extended to the decentralized setting in~\cite{centralized_secure_picod}, where a converse bound under the constraint of linear encoding showed a multiplicative gap of roughly three between the secure centralized and decentralized versions of PICOD.
This gap is strictly larger than the one between the centralized and decentralized versions of IC without security which equals two~\cite{embedded_ic}.

{\bf Contributions:}
We continue here the line of study initiated in~\cite{centralized_secure_picod}, where %In~\cite{centralized_secure_picod} 
we found: 
(i) there are infeasible cases %where there is no linear codes that can satisfy the decodability and security simultaneously.
%We observed that most of these cases are under the category that 
when $m$, the number of users, is odd, and %However, the 
(ii) the proposed achievable scheme %proposed in~\cite{centralized_secure_picod} did not apply to all $m$ and 
does not match the converse bounds (in a multiplicative or additive gap sense) in general. 
%The question for feasibility of general $m$ remained open in\cite{centralized_secure_picod}.
This paper completely answers the question posed in~\cite{centralized_secure_picod} by 
(i) providing a {complete} list of infeasible cases, which shows that most cases when $m$ is odd are actually feasible, and
(ii) showing achievable schemes for all feasible cases. 
We conclude that the new schemes achieve the converse bounds in~\cite{centralized_secure_picod} to within an {\it additive} constant gap.
% \begin{enumerate}
%     \item providing a full list of infeasible cases. The list shows that most of the odd $m$ cases are feasible;
%     \item showing achievable schemes for all feasible cases. The schemes achieves the converse bounds in~\cite{centralized_secure_picod} within an additive constant gap.
%     % , therefore shows that the converse bound is linearly optimal.
% \end{enumerate}

{\bf Paper Organization:}
The rest paper is organized as follows.
Section~\ref{sec:problem formulation} introduces the problem.
Section~\ref{sec:contribution} summarizes the main results in this paper.
In Section \ref{sec:infeasibility} we prove the infeasible cases. %mentioned in Theorem \ref{thm:infeaibility}. 
In Section~\ref{sec:intuition of achievability} we illustrate the main ideas for the achievable scheme by way of examples to provide the intuitions for the general scheme. 
Section~\ref{sec:summarize} concludes the paper.
The details of the proofs %for Theorem \ref{thm:constant_gap}
can be found in the Appendix.

% {\red
% maybe use the following 
% %
% $\ell^\star_{\textrm{IC}+\textrm{D}} / \ell^\star_{\textrm{IC}} \leq 2$~\cite{embedded_ic};
% $\ell^\star_{\textrm{IC}+\textrm{S}} / \ell^\star_{\textrm{IC}} \leq \infty$ because of infeasible casses;
% $\ell^\star_{\textrm{IC}+\textrm{D}+\textrm{S}} / \ell^\star_{\textrm{IC}+\textrm{S}} \leq ??$;
% %
% $\ell^\star_{\textrm{PICOD}+\textrm{D}} / \ell^\star_{\textrm{PICOD}} \leq 2$ from~\cite{embedded_ic} but often much less~\cite{decentralized_picod};
% $\ell^\star_{\textrm{PICOD}+\textrm{S}} / \ell^\star_{\textrm{PICOD}} \leq \infty$  because of infeasible casses;
% $\ell^\star_{\textrm{PICOD}+\textrm{D}+\textrm{S}} / \ell^\star_{\textrm{PICOD}+\textrm{S}} \leq 3$ this work;
% %

% }

\section{Problem formulation}
\label{sec:problem formulation}
            
We consider the $(m,s)$ secure decentralized PICOD problem with circular side information structure at the users.
The system consists of $m$ messages and $m$ users. 
The messages are vectors of length $\kappa\in\mathbb{N}$ of independent and uniformly distributed bits.
$\mathcal{W}:=\{w_1,\dots,w_m\}$ denotes the set of all messages. 
$W_A:=\{w_i, i\in A\}$ denotes the set of messages with indices in set $A$.
$\mathcal{U}:=\{u_1,\dots,u_m\}$ denotes the set of the users.
User $u_i$ has message $W_{A_i}$ as its side information, where $A_i =\{i,i-1\dots,i-s+1\}$. 
The entries in the side information set are intended modulo $m$.
The collection $\mathcal{A}=\{A_1,\dots,A_m\}$ is globally known to all users.
The coded symbols %codewords
are generated by the users based on their side information set. 
The encoding function at user $i$ is
\begin{align}
    x^{\kappa\ell_i} := \textbf{ENC}_i (W_{A_i}, \mathcal{A}), 
    \ i\in[m],
    \label{eq:encoding func}
\end{align}
where $\ell_i \in\mathbb{N}$ is the code length.
The overall transmission is represented by the vector $x^{\kappa\ell} = \{x^{\kappa\ell_1},\dots,x^{\kappa\ell_1}\}$ with
total normalized %number of transmissions 
length $\ell = \sum_{i\in[m]} \ell_i$.

Each user must correctly acquire %decode 
a message %$w_{d_i}$ 
that is not in its side information set. 
The decoding function at user $i$ is 
\begin{align}
    \widehat{w}_i:= \textbf{DEC}_i (x^{\kappa\ell}, W_{A_i}),
    \ i\in[m].
\end{align}
The decoding is correct %successful 
if $\widehat{w}_i= w_{d_i}$ for some $d_i\in [m]\setminus A_i$.

The security constraint requires user $i$ decodes no more than one message. %$w_{d_i}$.%u_i, \forall i\in[m]
Specifically, we have $I(w_j; x^{\kappa\ell}, W_{A_i})=0$ %\leq \ell\epsilon$ for 
for all $j\in[m]\setminus\{A_i\cup\{d_i\}\}$ and $i\in[m]$. %for arbitrary small $\epsilon$.

%For given $(m,s)$, 
The goal is to find the smallest $\ell$ %and the corresponding achievable scheme such
so that all users can correctly %successfully 
decode and the security constraints are satisfied.

\section{Main Contributions}
\label{sec:contribution}
%In this paper, we restrict the encoding functions to be linear.
For the $(m,s)$ secure decentralized PICOD with circular side information at the users, %and linear encoding function, 
the converse bound in~\cite{centralized_secure_picod} gives
\begin{align}
\label{eq:lower_bound}
    \ell^*\geq \begin{cases}
    \frac{m}{s}, & \frac{m}{m-s}\in \mathbb{Z}, \\
    \frac{3m}{2s}, & \frac{m}{m-s}\notin \mathbb{Z}, \textrm{linear encoding}, m > 2s, \\
2, & \frac{m}{m-s}\notin \mathbb{Z}, \textrm{linear encoding}, m < 2s.
    \end{cases}
\end{align}
% This bound is valid under the conditions:
% \begin{itemize}
%     \item Encoding function is invertible. That is, knowing the output and all but one inputs, we can find the unknown input. (Linear function is one of the examples).
%     \item $m\geq 2s$. This is the condition of argument that sending one message per transmission is not secure.
% \end{itemize}
% Because of the second condition, the lower bound here applies to only the region $m\geq 2s$. However, note that for the rest of the cases the lower bound is less than the constant 3. 
% Note that when 1-factor exists, the lower bound is strictly less than 2 and information theoretical optimal. 

In~\cite{centralized_secure_picod} we found several cases where the %secure decentralized PICOD
problem is infeasible, 
that is, no linear scheme exists such that every user can decode one and only one message outside its side information set. 
For such cases, the converse bound in~\eqref{eq:lower_bound} is not tight. 
In this paper, we first give %complete the
a list of infeasible cases and show scheme that  attain the
%Then for all feasible cases, we show that the 
converse bound in~\eqref{eq:lower_bound} to within an additive constant gap. 
Specifically, we have the following theorems.

\begin{thm}[\bf All infeasible cases]
    \label{thm:infeaibility}
    The secure $(m,s)$ decentralized PICOD with circular side information sets and linear encoding is infeasible if
    %it is not possible to satisfy all users while maintaining the security constraint only in the following cases:
    \begin{enumerate}
    \item $m\geq 2s+1$, $s=1$ or $2$;
    % \item $m\geq 5, s=2$;
    \item Odd $m\geq 7$, $s=3$ or $4$;
    % \item $m=7$, $s=4$, $m=7$;
    % \item Odd $m\geq 7$, $s=4$;
    \item Odd $m$, $s=m-2$.
   \end{enumerate}
\end{thm}

\begin{thm}[\bf Achievability to within an additive gap {of 7}]
\label{thm:constant_gap}
    For the $(m,s)$ secure decentralized PICOD with circular shift side information sets that are not described in Theorem \ref{thm:infeaibility},
    % when $ m/(m-s) \in \mathbb{Z} $ we have $\ell^{\star}_{\mathrm{it}}=m/s$.
    % Otherwise, the optimal code-length under linear encoding constraint is bounded as
    % , when $s\leq m/2$ we have  *I think the s<m/2 condition can be lifted*
    the following is attainable %number of transmissions suffices. 
    \begin{align}
\label{eq:upper_bound}
    \ell^*\leq \begin{cases}
    \frac{m}{s}, & \frac{m}{m-s}\in \mathbb{Z}, \\
    \frac{3m}{2s}+3, & \frac{m}{m-s}\notin \mathbb{Z}, \textrm{linear encoding}, m > 2s, \\
    2+5, & \frac{m}{m-s}\notin \mathbb{Z}, \textrm{linear encoding}, m < 2s.
    \end{cases}
\end{align}
    %In other words, the converse bounds in \eqref{eq:lower_bound} can be achieved within an additive constant gap. 
\end{thm}

% In the following sections we are going to show that the converse bounds in \eqref{eq:lower_bound} can be achieved within an additive constant gap. 
% For an achievability that has a constant number of transmissions, the optimality is within a constant additive gap. 

% Overall, the lower bound we have is
% \begin{align}
% \label{eq:lower_bound}
%     \ell^*\geq \begin{cases}
%     \frac{3m}{2s}, & m \geq 2s, \\
% 1, & m < 2s.
%     \end{cases}
% \end{align}

\section{Proof of Theorem~\ref{thm:infeaibility}} %Last infeasible case}
\label{sec:infeasibility}

% There are several cases where the secure decentralized PICOD are infeasible. 
%Infeasible cases in the centralized secure PICOD setting are also infeasible in the decentralized setting.
%Therefore, 
The two infeasible cases in the centralized secure PICOD setting in~\cite{private_picod}, namely
(i) Odd $m$, $s=1$, and (ii) Odd $m$, $s=m-2$,
% \begin{enumerate}
%     \item Odd $m$, $s=1$,
%     \item Odd $m$, $s=m-2$,
% \end{enumerate}
% 1) odd $m$, $s=1$; 2) odd $m$, $s=m-2$, 
are infeasible also in the decentralized setting.
% Compared with the centralized secure PICOD discussed in~\cite{private_picod}, there are a few more cases that becomes infeasible in the decentralized setting. 
In~\cite{centralized_secure_picod} we showed the infeasiblility in the decentralized setting of
(iii) Even $m\geq 3$, $s=1$,
(iv) $m\geq 5$, $s=2$,
(v) Odd $m$, $s=3$.
% \begin{enumerate}
%     \item Even $m\geq 3$, $s=1$,
%     \item $m\geq 5$, $s=2$,
%     \item Odd $m$, $s=3$.
%     % \item Odd $m$, $s=3$.
% \end{enumerate}
% We complete the list of infeasible cases by showing the last infeasible case: odd $m\geq 7$, $s=4$.
In the rest of this section %, we complete the list of infeasible cases by showing
we prove the last infeasible case (vi) odd $m\geq 7$, $s=4$, and we split it into two cases.
    \subsection{Case Odd $m\geq 9$, $s=4$}

    Assume we have an achievable scheme that satisfies all %users and security
     constraints.
    Since here $m>2s$, from~\cite[Proposition~1]{private_picod}, one transmission involves at least 2 messages. 
    {
    The number of messages in one transmission is either 2 or 3, since involving $s=4$ consecutive messages in one transmission is insecure as shown in~\cite{centralized_secure_picod}.
    For a linear code, the user can decode its desired message if there exists one linear combination of the messages such that all the messages but its desired message that are involved in the linear combination are in the user's side information set. 
    Therefore, if two transmissions do not satisfy one user and involve no messages in common, the linear combination of these two transmissions will not satisfy the user. 
    Each transmission involving $2$ or $3$ consecutive messages satisfies $2$ users.
    Each transmission involving $2$ nonconsecutive messages satisfies $4$ users.
    Thus if no transmissions have common messages the total number of satisfied users are even, which contradicts to the condition that $m$ is odd.
    }
    % Note that 
    % If no two transmissions involve a common message, the total number of users satisfied is the sum of the number of satisfied users by each transmission.
    % Note that each transmission satisfies an even number of users \dt{WHY?}; if $m$ is odd, this generates a contradiction.
    Therefore, there must exists two transmissions that have messages in common.
    % Note that involving 4 different messages in one transmission is insecure. 
    We consider the following three sub-cases.
    
    \paragraph{Both transmissions involve 2 messages}
    {There exists a user that can decode the common message by one transmission then decode another message by the other transmission.}
    Therefore the user can decode two messages and violates the security constrain.

    \paragraph{One transmission involves 2 messages and the other 3 messages}
    \label{para:2 and 3 msg}
    Let $g(.)$ denote a linear combination of its argument.
    We have the following cases: %, which violate the security constrain:
        \begin{enumerate}
            \item $g_1(W_{4,5})$ and $g_2(W_{4,5,6})$: the user with side information $W_{1,2,3,4}$ can decode $W_{5,6}$, which is insecure\label{item:case 1};
            \item $g_1(W_{4,5})$ and $g_2(W_{5,6,7})$: the user with side information $W_{6,7,8,9}$ can decode $W_{4,5}$, which is insecure;
            \item $g_1(W_{4,6})$ and $g_2(W_{4,5,6})$: the user with side information $W_{1,2,3,4}$ can decode $W_{5,6}$, which is insecure;
            \item $g_1(W_{4,6})$ and $g_2(W_{5,6,7})$: the user with side information $W_{6,7,8,9}$ can decode $W_{4,5}$, which is insecure;
            \item $g_1(W_{3,5})$ and $g_2(W_{5,6,7})$: the user with side information $W_{6,7,8,9}$ can decode $W_{3,5}$, which is insecure.
        \end{enumerate}
    %therefore, the security constrain is violated.

    \paragraph{Both transmissions involve 3 consecutive messages}
    \label{para:3 and 3 msg}
    We have the following cases:
    \begin{enumerate}
            \item $g_1(W_{4,5,6})$ and $g_2(W_{4,5,6})$: 
            a linear combination of these two transmissions can generate a linear combination of $W_{4,5}$. {This case is insecure as shown in case \ref{item:case 1} of \ref{para:2 and 3 msg}}; %and the code becomes insecure.
            \item $g_1(W_{4,5,6})$ and $g_2(W_{5,6,7})$:
            the user with side information $W_{2,3,4,5}$ can decode $W_{6,7}$, which is insecure;
            \item $g_1(W_{4,5,6})$ and $g_2(W_{6,7,8})$:  
            a linear combination of these two transmissions can generate a linear combination of $W_{4,5,6,7,8}$.
             % with $w_{6}$ possibly absent. 
            %The new linear combination 
            % \dt{WHY? HOW MANY WERE SATISFIED SO FAR? I INCLUDED THE REST IN THIS BULLET POINT AS I DID NOT SEE HOW IT FIT OTHERWISE}. 
            % The user that can decode by linear combination can decode based on either the linear combination of $W_{4,5,6}$ or $W_{6,7,8}$. 
            This is the only case that does not violate the security constraint.
            However, this case does not allow any new users to decode. 
            % We conclude that if the messages in the codewords are reused in two transmissions, it has to be this case, i.e., the two transmissions are linear combinations of 3 messages that share one message. 
%%%%        
%%%%
%%%%    % We now consider the number users satisfied by the transmissions. 
%%%%    % If two transmission do not have common messages, their linear combinations do not help. 
%%%%    % The total  number of satisfied users is the sum of the number of the satisfied users of each transmission.
%%%%    % If two transmissions have common message, 
%%%%    Therefore, even with possible common messages, the transmissions' linear combinations do not help decoding more messages. 
%%%%    % it can only be the last case,  where the linear combinations of the transmissions does not allow the new user to decode. 
    Therefore, the number of satisfied users  is  the the sum of the number of users satisfied by each transmission, which even. 
    This contradicts the condition that $m$ is odd. 
    \end{enumerate} 

    % The number of users satisfied by each transmission is either 4 or 2, which are even.
    % We notice that when $s=4$, each transmission satisfies even number of users. 
    % Therefore, the total number of users satisfied is even. 
    % However, $m$ is odd,  which is a contradiction to the assumption that all users are satisfied. 
    % Therefore we are always left with one message that can not be satisfied under the security constraint. 
    This concludes the proof that the case odd $m\geq 9$ and $s=4$ is infeasible.

    \subsection{Case $m=7$, $s=4$}

    Assume we have an achievable scheme that satisfies all %users and the security 
    the constraints.
    If all the transmissions are linear combinations of at least 2 messages, the argument for the case $s=4$, odd $m\geq 9$ holds and thus the case is infeasible. 
    % The difference in this case is that $s\geq m/2$. 
    % Therefore,  the argument that sending one message in one transmission is  always insecure does not hold.
    Therefore, it is enough to show that the case $s=4, m=7$ is insecure if there is one transmission that involves only one message. 
    Without loss of generality, assume one transmission is a linear function of $w_{4}$. 
    Users $u_1,u_2,u_3$ will decode $w_{4}$ since they do not have it in their side information set. 
    {User $u_3$ have all the side information of user $u_4$ after decoding. 
    Therefore, the desired message of user $u_4$ needs to be inside the side information of $u_3$. 
    The desired message of $u_4$ can only be $w_7$.
    % will desire $w_{7}$ 
    The the argument applies to $u_7$ and its desired message can only be $w_1$
    }
        % 5 users' desired messages have been fixed. 
    Thus, only the desired message of user $u_{5}$ and user $u_{6}$ are not fixed yet.
    Consider the following cases for the desired message of user $u_{5}$. 
    % There are 3 possible desired messages for user $u_{5}$. 

    \paragraph{$d_5=1$}
    User $u_5$ can mimic $u_4$ and decode $w_7$. 
    % Insecure.
    \paragraph{$d_5=7$}
    There must exists a linear combination of codewords that is a linear combination of messages $W_{2,3,5,7}$ {by the decoding condition of the linear index code}, where $w_{7}$ has non-zero coefficient. 
    Let $k\in \{2,3,5\}$ be the largest index in $\{2,3,5\}$ so that $w_k$ has non-zero coefficient in said linear combination. 
    User $u_{k-1}$ can decode $w_k$ when $k=2,3$, user $u_3$ can decode $w_5$ when $k=5$. 
    Since these users already have other desired messages, they now decode more than one messages. 
    % Insecure.
    
    \paragraph{$d_5=6$}
    There must exist a linear combination of codewords that is a linear combination of messages $W_{2,3,5,6}$ {by the decoding condition of the linear index code}, where the coefficient of $w_{6}$ is non-zero.
    Therefore, there exists one transmission involving $w_6$. We consider all possible linear combinations that involve $w_6$.
    \begin{itemize}
        \item Linear combination of $w_6$, or $W_{4,6}$, or $W_{1,6}$, or $W_{6,7}$, or $W_{1,6,7}$. 
        It will allow user $u_3$ to decode parts of $w_6$. 
        % Since $d_3=4\neq 6$ this case is insecure.
        \item Linear combination of $W_{5,6}$ or $W_{4,5,6}$. 
        It will allow user $u_2$ to decode of $w_5$ {since user $u_2$ has $w_4$ as the desired message}. 
        % Since $d_2=4\neq 5$ this case is insecure.
    \end{itemize}
    Therefore, the case $s=4, m=7$ is infeasible.

    Overall, we conclude that the case $s=4$, odd $m\geq 7$ is infeasible.
    This completes the proof of the infeasible case list in Theorem~\ref{thm:infeaibility}.

\section{Proof of Theorem~\ref{thm:constant_gap}} %Achieving the converse bounds}
\label{sec:intuition of achievability}

In this section, we give examples to demonstrate the key ideas in our achievable schemes.
% RESERVED FOR ITW VERSION !!! 
% \dt{The details on the general case can be found in the longer version of this paper at~\cite{ourarxiv}.} 
% The converse bound in~\eqref{eq:upper_bound} includes three regimes, which we shall address separately.  
The details on the general case can be found in Appendices.
% The converse bound in~\eqref{eq:upper_bound} includes three regimes, which we shall address separately.  

%\begin{enumerate}
%   \item $\ell^*=\frac{m}{s}$ when $\frac{m}{m-s}\in\mathbb{Z}$.
%   \item $\ell\leq \frac{3m}{2s}+3$ when $\frac{m}{m-s}\notin\mathbb{Z}, m>2s$.
%   \item $\ell \leq 9$ when $\frac{m}{m-s}\notin\mathbb{Z}, s<m<2s$.
%\end{enumerate}
%
%We address the schemes for these three regimes separately.

\subsection{Case $\frac{m}{m-s}\in\mathbb{Z}$: $\ell^*=\frac{m}{s}$}
\label{sub:1-factor exist}

%This is the case where 
The information theoretical %optimal number of transmissions 
converse was derived in~\cite{centralized_secure_picod} for the centralized case.
The achievable scheme is the one for the decentralized PICOD without security constraint discussed in~\cite{decentralized_picod}, which satisfies the security constraint because for each user, among all the messages that are involved in the encoding function, there is one and only one message that is not in its side information set. 
%In~\cite{decentralized_picod} the scheme has been shown to be information theoretically optimal.
Therefore, the scheme is also information theoretically optimal with an additional security constraint. 

% When $\frac{m}{m-s}\in\mathbb{Z}$, the achievable scheme in~\cite{decentralized_picod}, which is for the decentralized PICOD without security constraint, also satisfies the security constraint. 
% This is because among all messages involved in the encoding, for each user, there is one and only one message that is not in the side information set of the user. 
% Therefore, no user is able to decode any information from the messages that are not its desired message and the defined security constraint is thus satisfied.
% The proposed achievable scheme is information theoretically optimal.
% % The number of transmissions is $\ell^*_{it}=\frac{p}{p-1}$ where $p$ is the size of the 1-factor.
% % Since there exists only one 1-factor because the NTH is regular in this case, we conclude that the optimal number of transmissions is $\ell^*_{it}=\frac{m/(m-s)}{m/(m-s)-1}=\frac{m}{s}$.
% For this case, the number of transmissions is no greater than 2. 

\subsection{Case $\frac{m}{m-s}\notin\mathbb{Z}, m>2s$: $\ell\leq \frac{3m}{2s}+3$}
\label{sub:1-factor notexist, m>2s}

The converse bound in this regime is $\frac{3m}{2s}$~\cite{private_picod}.
Therefore, in order to have an achievable scheme that is optimal to within an additive constant gap, we aim to satisfy {on average} $\frac{2s}{3}$ users in one transmission.

It has been shown that, for the cases where $\frac{m}{2s}\in \mathbb{Z}$, an optimal secure centralized scheme is  $\{w_{1+2sk}+w_{2+2sk}, w_{3+2sk}+w_{s-2+2sk}, w_{s-3+2sk}+w_{s-4+2sk}\}, k\in\{0,1,\dots,\frac{m}{2s}-1\}$, for a total $\frac{3m}{2s}$ transmissions~\cite{centralized_secure_picod}. % and satisfies all $m$ users securely
% This scheme satisfies $2s$ users with every three transmissions while keeping the security constraint satisfied. 
% On average, each transmission satisfies $2s/3$ users. 
% It therefore meets the converse bound.
In this section, by ways of examples, we show how this scheme can be extended %such that the converse bound can be achieved within a constant gap for
to all feasible cases in the regime $\frac{m}{m-s}\notin\mathbb{Z}, m>2s$.

To make the exposition easier, we shall use the case $m=26$ in the following and represent the scheme in a ``matrix'' form figures.
In the figure, a row with \textbf{X}'s represents a transmission that is a linear combination of the messages marked by the \textbf{X}.
%The \textbf{X}s indicate the indices of the messages that are involved in the linear combination.
A row with \textbf{U}'s shows the users that are satisfied by the transmission shown by the row right above. 
The user represented by the \textbf{U} in position $i$ is the user with side information set $W_{A_i} =\{w_{i-s+1},\dots,w_{i}\}$. %, where all the subscripts are in modulo $m$. A_i

% Generally speaking, our scheme groups the users into the $k$ groups of size $2s$, where $k\in\mathbb{Z}$ and $0\geq m-2ks<2s$. 
% If $\frac{m}{2s}\notin\mathbb{Z}$, there are $m-2ks \neq 0$ users outside the group.
% % We use scheme that is proposed in~\cite{centralized_secure_picod}  to satisfy the users in the groups. 
% % Each group requires three transmissions. 
% $3k$ transmissions are used to satisfy the users in the groups.
% For the $m-2ks$ users outside the groups, we use a constant $c$ number of transmissions.
% The scheme thus requires $3k+c$ transmissions in total. 
% Note that $3k+c\leq \frac{3m}{2s}+c$, this scheme achieves the converse bound within a constant gap. 
% The details of the schemes and rigorous can be found in Appendix~\ref{sec: m>2s}.

% \subsubsection{Examples}
% \label{ssub:examples for m>2s}

\paragraph{Case $m=26, s=6$}
\label{para: ex:m=26,s=6}

The %transmissions and the users that are satisfied by each transmission are 
scheme is illustrated in Fig.~\ref{fig:ex:m=26,s=6}.
We have two groups of $2s=12$ users each. % of size . 
Six transmissions are used to satisfy the 24 users in these groups. 
The transmissions are $w_{1}+w_{2}, w_{3}+w_{7}, w_{8}+w_{9}, w_{13}+w_{14}, w_{15}+w_{19}, w_{20}+w_{21}$.
% , shown as the rows in Fig.~\ref{fig:ex:m=26,s=6}.
The two remaining users % and we use one transmission to satisfy the remaining users.
are satisfied by the last transmission $w_{22}+w_{23}+w_{24}+w_{25}+w_{26}$.
% , shown as the row in Fig.~\ref{fig:ex:m=26,s=6}.
The total number of transmissions is $\ell=7$.

% This scheme satisfies the security constraint because the messages involved in different transmissions are all different, i.e., no message has been used in more than one transmission. 
% Therefore, the linear combination of the transmissions does not help decoding. 
% This scheme guarantees that each user can decode one message from only one transmission.
% The users thus can not decode more than one message from the received codewords and their side information.

% This is the case where the remaining users can be securely satisfied by one transmission.

\begin{figure}%[ht]
    \centering
    \includegraphics[width=0.9\columnwidth]{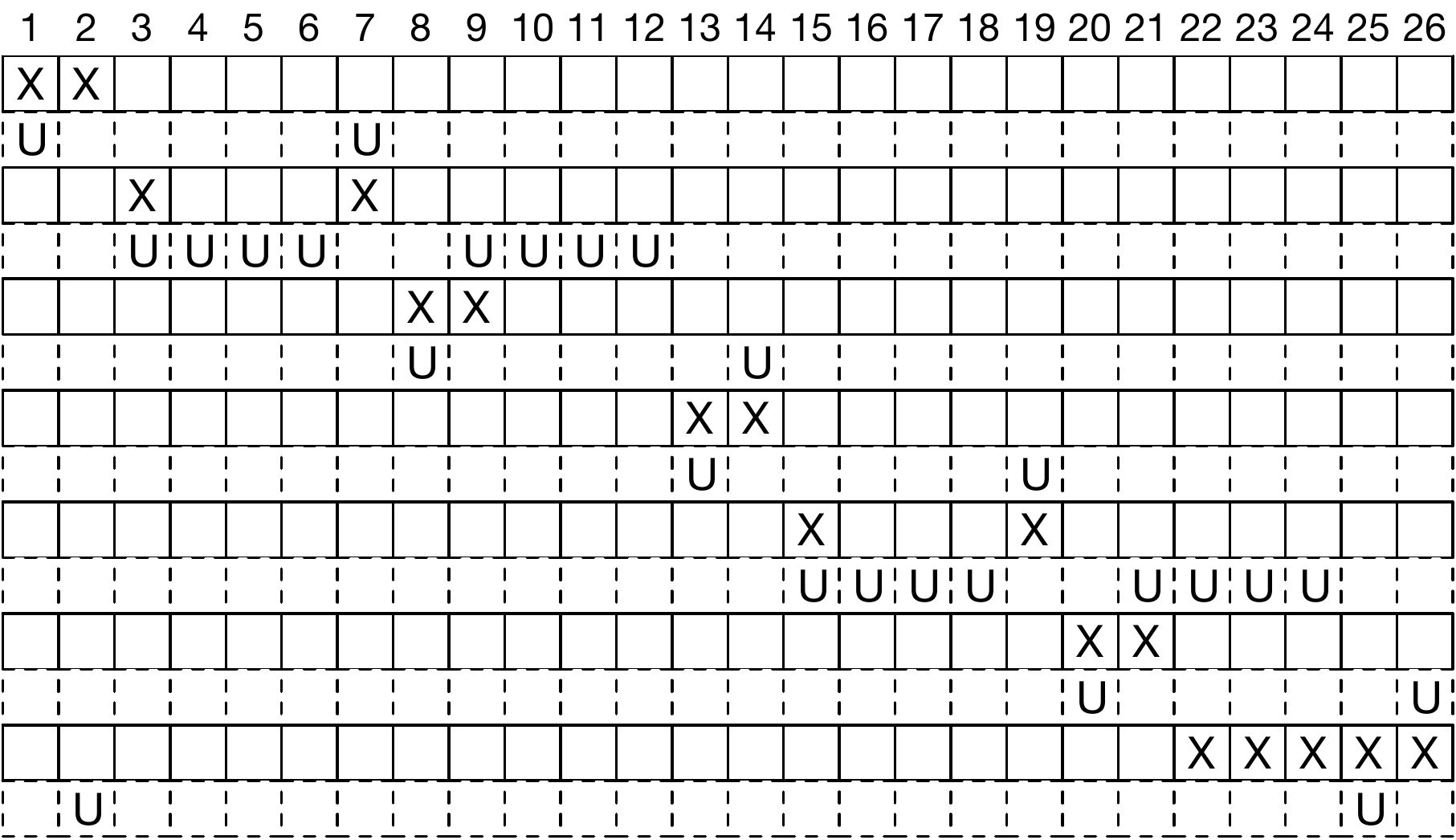}
%     \begin{tikzpicture}
% \matrix (m) [matrix of nodes,
%              nodes={draw,minimum size=1em, outer sep=0pt,inner sep=0,line
%              width=0.5pt,append after command={\pgfextra{\draw 
%              ($(\tikzlastnode.north west)+(-0.5em,+0.5em)$)
%              rectangle ($(\tikzlastnode.south east)+(0.5em,-0.5em)$);}}},
%              nodes in empty cells,column sep=-0.5pt, row sep=-0.5pt
%               ]
% {
%    $w_1$  &  &   \\
%     3 & 5 & 6  \\
%     7 & 8 & 9  \\
% };
% \end{tikzpicture}
    \caption{Achievable scheme for $m=26, s=6$.}
    \label{fig:ex:m=26,s=6}
\end{figure}

\paragraph{Case $m=26, s=10$}
\label{para: ex:m=26,s=10}

We have one group of $2s=20$ users, and $6$ remaining users.
The transmissions and the users that are satisfied by each transmission are illustrated in Fig.~\ref{fig:ex:m=26,s=10}.
Three transmissions satisfy the $20$ users' group. 
The transmissions are $w_{1}+w_{2}, w_{3}+w_{12}, w_{13}+w_{14}$.
For the remaining users, we use two transmissions
$w_{15}+w_{17}+w_{18}+w_{19}+w_{20}+w_{21}+w_{22}$ and $w_{18}+w_{19}+w_{20}+w_{21}+w_{22}+w_{23}+w_{25}$,
Each satisfying three of the remaining users. 
The total number of transmissions is $\ell=5$.
% Only the last two transmissions involve messages in common. 
% One can check that the linear combination of these two messages does not allow any users to decode any messages other than their desired message.
% Therefore, the scheme satisfies the security constraint.

% This is the case where the scheme needs two transmissions to satisfy the remaining users.

\begin{figure}%[ht]
    \centering
    \includegraphics[width=0.9\columnwidth]{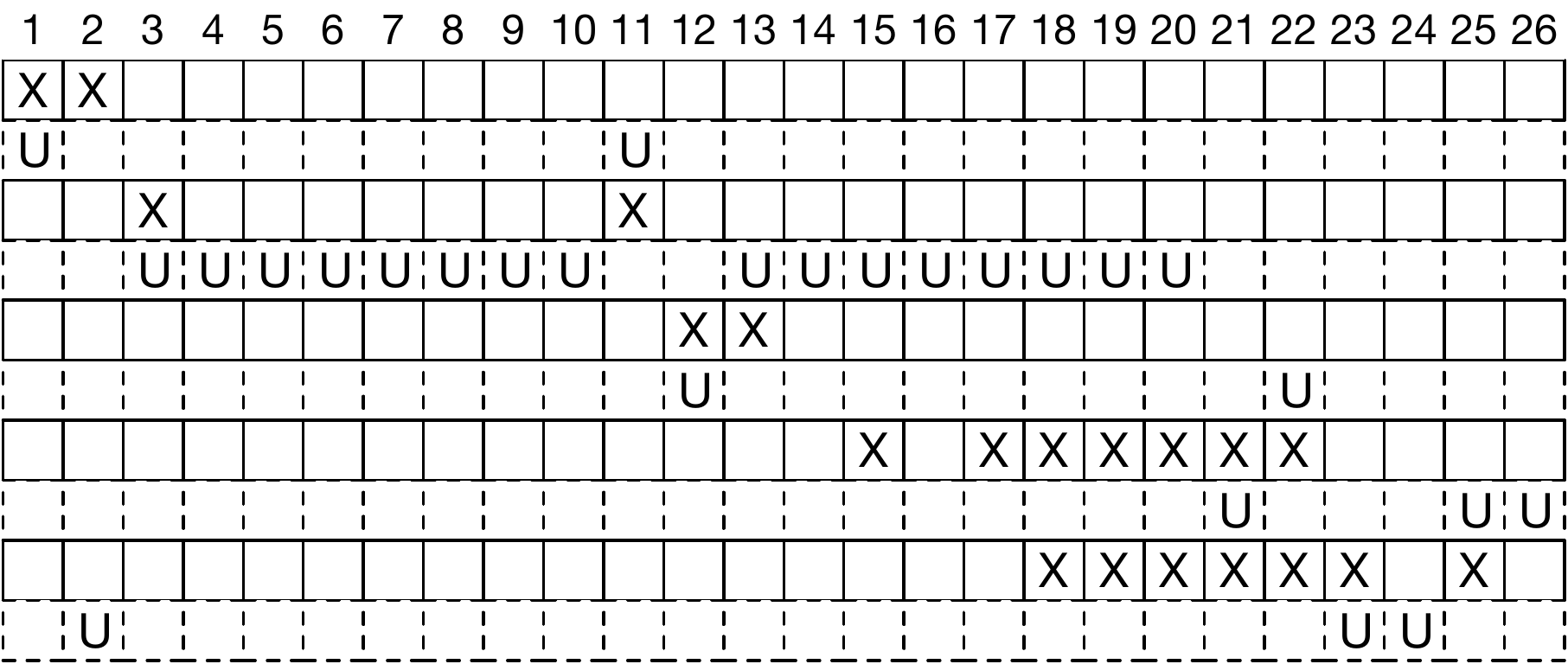}
    \caption{Achievable scheme for $m=26, s=10$.}
    \label{fig:ex:m=26,s=10}
\end{figure}

% %!! DELETING ONE EXAMPLE
% \paragraph{Case $m=26, s=7$}
% \label{para: ex:m=26,s=7}

% We have one user group of size $2s=14$ and $12$ remaining users.
% The transmissions and the users that are satisfied by each transmission are illustrated in Fig.~\ref{fig:ex:m=26,s=7}.
% Three transmission are used to satisfy the $14$ users in the group.
% The transmissions are $w_{1}+w_{2}, w_{3}+w_{8}, w_{9}+w_{10}$.
% We use three transmissions for the remaining users:
% $w_{17}+w_{20}, w_{13}+w_{15}+w_{16},  w_{21}+w_{22}+w_{24}$.
% The transmission $w_{17}+w_{20}$ satisfies 6 users, leaves the other 6 users unsatisfied. 
% Then the remaining two transmissions each satisfies three users each. 
% The total number of transmissions is $\ell=6$.
% % Following the same argument in Section~\ref{para: ex:m=26,s=6}, the security constraint is satisfied.

% % In this case the scheme needs three transmissions to satisfy the remaining users.

% \begin{figure}%[ht]
%     \centering
%     % \includegraphics[width=0.8\columnwidth]{scheme examples/ex_m26_s_7.png}
%     \includegraphics[width=0.8\columnwidth]{FIG/main_examples/s=7,m=26.pdf}
%     \caption{Achievable scheme for $m=26, s=7$.}
%     \label{fig:ex:m=26,s=7}
% \end{figure}

\begin{rem}
\label{rem:m>2s}
    The key ideas here are as follows.
    In the regime $m>2s$ we use the scheme that satisfies $2s$ users by using three transmissions.
    In the first step, %stage of the transmission, 
    we group the users into disjoint groups of size $2s$ and satisfy the users in each group % optimally -- the number of users satisfied by each transmission is
    with $3$ transmissions.
    %We are left with some users that are outside the groups. 
    In the second step, %stage, 
    we satisfy the remaining users,
    %Since the number of these users 
    which are less than $2s$.
    % For the case where the number of remaining users is not 0,  the converse bound shows that the optimal scheme needs no less than 
    % Since the number 
    {Our scheme guarantees that the number of transmissions needed to satisfy the remaining users is a constant that does not grow with the system parameter $(m, s)$.}
    % The number of transmissions needed to satisfy the remaining users is upper bounded by a constant, which does not grow with the system parameter $(m, s)$ \dt{WHY?}.
    Therefore, for the feasible cases, the proposed scheme can achieve the converse bound to within an additive constant gap that equals the number of transmissions in the second step. 
\end{rem}

% \subsubsection{Summarize}
% \label{ssub:sumarize for m>2s}

\subsection{Case $\frac{m}{m-s}\notin\mathbb{Z}, s<m<2s$: $\ell \leq 9$}
\label{sub:1-factor notexist, m<2s}

In this regime the scheme in~\cite{centralized_secure_picod} does not work because the number of users is $m<2s$, thus no group of size $2s$ users can be formed. 
%To achieve the converse bound within a constant gap, 
Here we aim for satisfying all users with a constant number of transmissions that does 
%That is, the number of transmissions needed to satisfy all users should 
not grow with the system parameters $(m,s)$. 
%Our achievable scheme further splits the case $\frac{m}{m-s}\notin\mathbb{Z}, s<m<2s$ into two sub-cases:
%1) $\frac{m}{m-s}\notin\mathbb{Z}, s<m\leq \frac{3s}{2}$,
%2) $\frac{m}{m-s}\notin\mathbb{Z}, \frac{3s}{2}<m<2s$, 
We treat two sub-cases separately.

% ======================
% ======================
% ======================
\subsubsection{Subcase $\frac{m}{m-s}\notin\mathbb{Z}, s<m\leq \frac{3s}{2}$}
\label{ssub: m<3s/2}

In this case we consider the complement of the side information set of every user, which is of size $m-s$.
% Since $m<3s/2$ we have $m-s<s/2$.
The proposed scheme guarantees that among every consecutive $m-s$ messages, the codewords contain one and only one message that is linearly independent of the remaining $m-s-1$ messages.
 % that %. All the rest $m-s-1$ messages 
% are aligned in one \dt{linear  dimension?!?!?}. 
%Thus the decodability and security constraint are achieved by this scheme.
In the following, we provide examples to demonstrate the scheme.
The detailed proof can be found in Appendix~\ref{sec:s<m<3s/2}.

\paragraph{Case $m=26, s=20$}
\label{para: ex:m=26,s=20}

The codewords contain two parts: 
\begin{itemize}
    \item Part A with the structure $w_{i+1}+\ldots+w_{i+(m-s-1)}$, $w_{m-s}+w_{i+2(m-s-1)}$. It takes totally $2k_{1}(m-s-1)$ consecutive messages, where $k_{1}\geq 2, k_1\in\mathbb{Z}$.
    \item Part B with the structure $w_{j}, w_{j+1}+\ldots+w_{j+(m-s)}$. It takes totally $k_2(m-s)$ consecutive messages, where $k_2\geq 0, k_2\in\mathbb{Z}$.
\end{itemize}
% 1) multiples of $2(m-s-1)$; 2) multiple of $m-s$.  
Therefore, this scheme works for the case $m=k_{1}2(m-s-1)+k_{2}(m-s)$, where $k_{1}\geq 2, k_2\geq 0, k_1,k_2\in\mathbb{Z}$.
% , the scheme satisfies all users while meets the security constraint. 

In the case of $s=20, m=26=2(10)+6$, the proposed scheme takes $4$ transmissions:
1) $w_1+w_2+w_3+w_4+w_5+w_{21}+w_{22}+w_{23}+w_{24}+w_{25}$;
2) $w_6+w_10+w_{26}$;
3) $w_{11}+w_{12}+w_{13}+w_{14}+w_{15}$;
4) $w_{16}+w_{20}$,
as shown in Fig.~\ref{fig:ex:m=26,s=20}.

% One can check for all consecutive $6$ messages, there exists one and only one message that is linearly independent from the other 5 messages.

\begin{figure}%[ht]
    \centering
    \includegraphics[width=0.9\columnwidth]{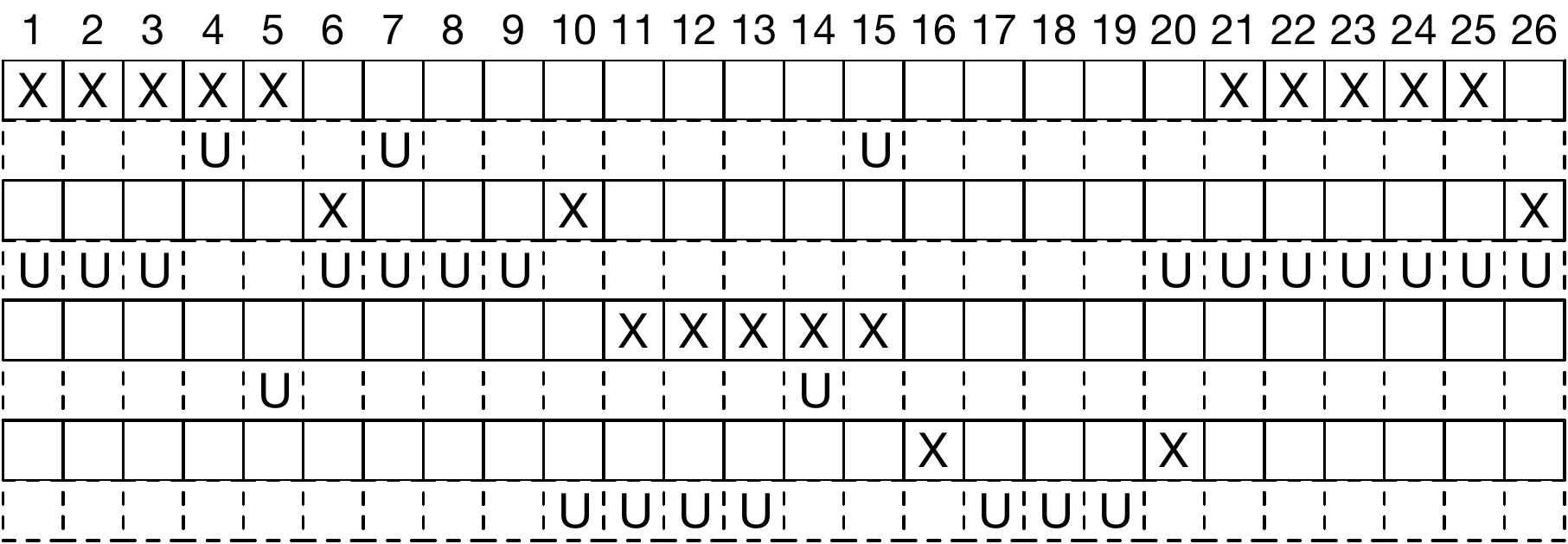}
    \caption{Achievable scheme for $m=26, s=20$.}
    \label{fig:ex:m=26,s=20}
\end{figure}

For the cases where $m\neq k_{1}2(m-s-1)+k_{2}(m-s), \forall k_1\geq 2, k_2\geq 0, k_1,k_2\in\mathbb{Z}$, we propose the following examples to show how the scheme can be modified.
 % to satisfy the cases.

\paragraph{Case $m=26, s=19$}
\label{para: ex:m=26,s=21}

In this case the proposed scheme takes $4$ transmissions: 
1) $w_5$;
2) $w_2+w_3+w_{4}+w_{6}+w_{11}+w_{22}$;
3) $w_{12}+w_{13}+w_{14}+w_{15}$;
4) $w_{18}+w_{21}+w_{24}$;
as shown in Fig.~\ref{fig:ex:m=26,s=19}.
One can check that for every $7$ consecutive messages, there exists one and only one message that is linearly independent of the other $6$ messages. 
The scheme can be seen as a modified version of the scheme in Section~\ref{para: ex:m=26,s=20}:
the part from $w_{6}$ to $w_{17}$ is a modified Part~A;
the part from $w_{25}$ to $w_{5}$ is a modified Part~B; and
the rest is a new structure that combines these two modified pieces.

\begin{figure}%[ht]
    \centering
    \includegraphics[width=0.9\columnwidth]{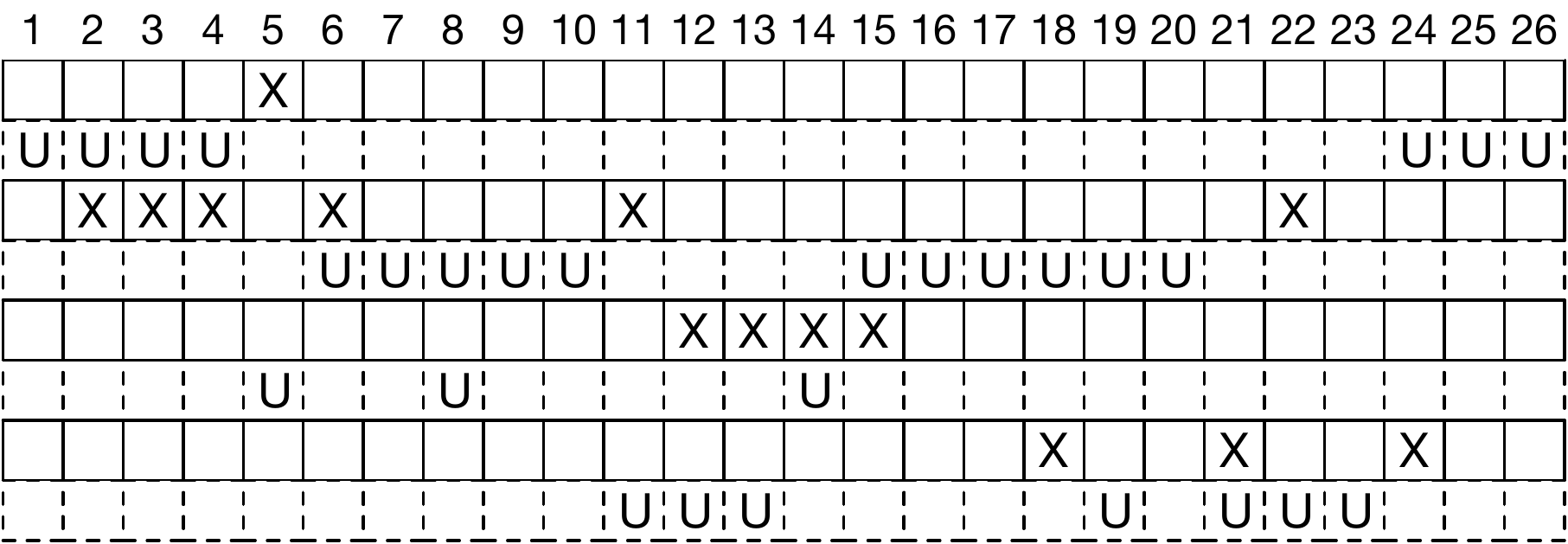}
    \caption{Achievable scheme for $m=26, s=19$.}
    \label{fig:ex:m=26,s=19}
\end{figure}

%!! DELETING ONE EXAMPLE 
\paragraph{Case $m=26, s=18$}
\label{para: ex:m=26,s=18}

In this case the proposed scheme takes $4$ transmissions: 
1) $w_1+w_4+w_5+w_6+w_7$;
2) $w_8+w_{13}$;
3) $w_{15}+w_{16}+w_{17}+w_{18}+w_{19}$;
4) $w_{22}+w_{26}$,
as shown in Fig.~\ref{fig:ex:m=26,s=18}.
One can check for every $8$ consecutive messages, there exists one and only one message that is linear independent from the other $7$ messages. 
The scheme can be seen as a modified version of the scheme used in Section~\ref{para: ex:m=26,s=20}:
the part from $w_{1}$ to $w_{14}$ is a modified part A;
the part from $w_{15}$ to $w_{26}$ is a modified and shrunk part A.

\begin{figure}%[ht]
    \centering
    \includegraphics[width=0.9\columnwidth]{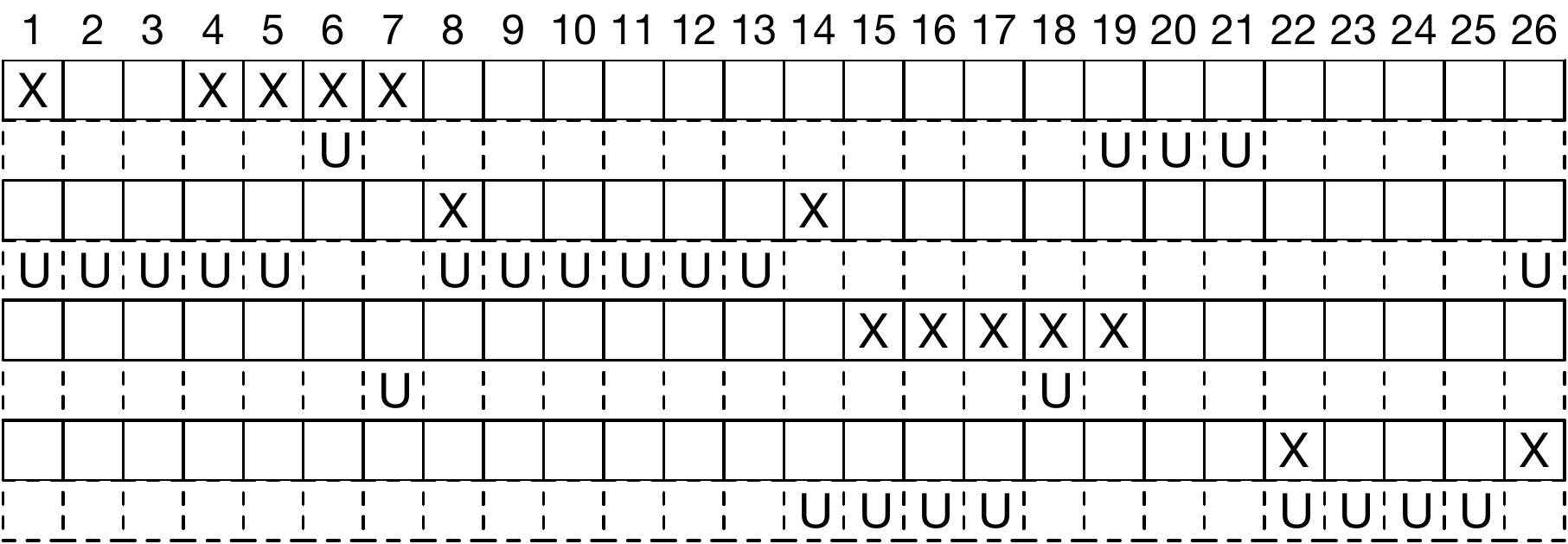}
    \caption{Achievable scheme for $m=26, s=18$.}
    \label{fig:ex:m=26,s=18}
\end{figure}

\begin{rem}
\label{rem:m<3s/2}
    The key ideas here are as follows.
    In this regime our proposed scheme is designed such that 
    % satisfies both decodability and security by design the codewords 
    every consecutive $m-s$ messages have one and only one messages that is linearly independent of the rest. 
    The scheme treats the users is a ``sliding window'' fashion, where the window is of size $m-s$. %and design the codewords accordingly. 
    % The idea is different from the idea of grouping users and satisfying users per group. 
    % The reason the ``grouping'' method is not used here is that in this case $m<2s$ thus no group of size $2s$ can be made. 
    % The ``sliding window'' idea is preferred when $m-s<s/2<m/s$ because
    {The scheme for case $m>2s$ can not be extended straightforwardly in this case because to satisfy users less than $2s$ the scheme has transmissions that takes more than $2s$ consecutive messages. 
    This is not an issue in the case $m>2s$. 
    But when $m<2s$ this means the transmissions will have messages in common as the indices of the messages are in modulo $m$. 
    The common messages among the transmissions will cause insecurity as it could allow one user to decode more than one message.
    The security constraint does not hold anymore. 
    % When $m-s<s/2<m/2$ the size of the ``window'' is smaller than $m/2$. 
    % There exist disjoint "windows" that do not share messages in common. 
    The sliding window idea does not have such issues when considering about the security constraint. 
    }
    % \dt{less likely to be intersected with each other in this case , making the designing of the codewords easier to guarantee the decodability and security at the same time.??? THIS DOES NOT MAKE MUCH SENSE TO ME???}
\end{rem}

% ======================
% ======================
% ======================
\subsubsection{Subcase $\frac{m}{m-s}\notin\mathbb{Z}, \frac{3s}{2}<m<2s$}
\label{ssub: 3s/2<m<2}

Here we propose the scheme that can be thought of as a combination of the ideas we presented in Sections~\ref{sub:1-factor notexist, m>2s} and~\ref{ssub: m<3s/2}.
That is, we first use a finite number of transmissions to satisfy a finite number of users with the scheme in~\cite{centralized_secure_picod}.
Then, we use a scheme similar to the one proposed for the case $m>2s$ in Sections~\ref{sub:1-factor notexist, m>2s} for the remaining users.
% group the rest of the users  into groups and satisfy these users using the scheme similar to the ones 
The total number of transmissions is thus finite and the scheme achieves the converse bound to within a constant gap.
The following example illustrates the proposed scheme.
The detailed proof of the scheme can be found in Appendix~\ref{sec:3s/2<m<2s}.
 % in \cite{???}.

\paragraph{Case $m=26, s=16$}
\label{para: ex:m=26,s=16}
The first two transmissions are:
1) $w_1+\dots+w_{15}$;
2) $w_3+w_{17}$.
Four users are satisfied by these two transmissions.
The remaining users are grouped into $6$ groups and are satisfied by $3$ transmissions, where each transmission aims to satisfy $2$ groups.
The transmissions are:
3) $w_{20}+w_{24}$;
4) $w_8+w_{14}+\dots+w_{19}$;
5) $w_1+\dots+4+w_{10}+w_{25}+w_{26}$.
The transmissions and the corresponding satisfied users are shown in Fig.~\ref{fig:ex:m=26,s=16}.
The scheme uses $5$ transmissions to satisfy all users.

% One can check that no users can decode more than one message by its side information and the received codewords. Therefore the security constraint is satisfied as well.

\begin{figure}%[ht]
    \centering
    \includegraphics[width=0.9\columnwidth]{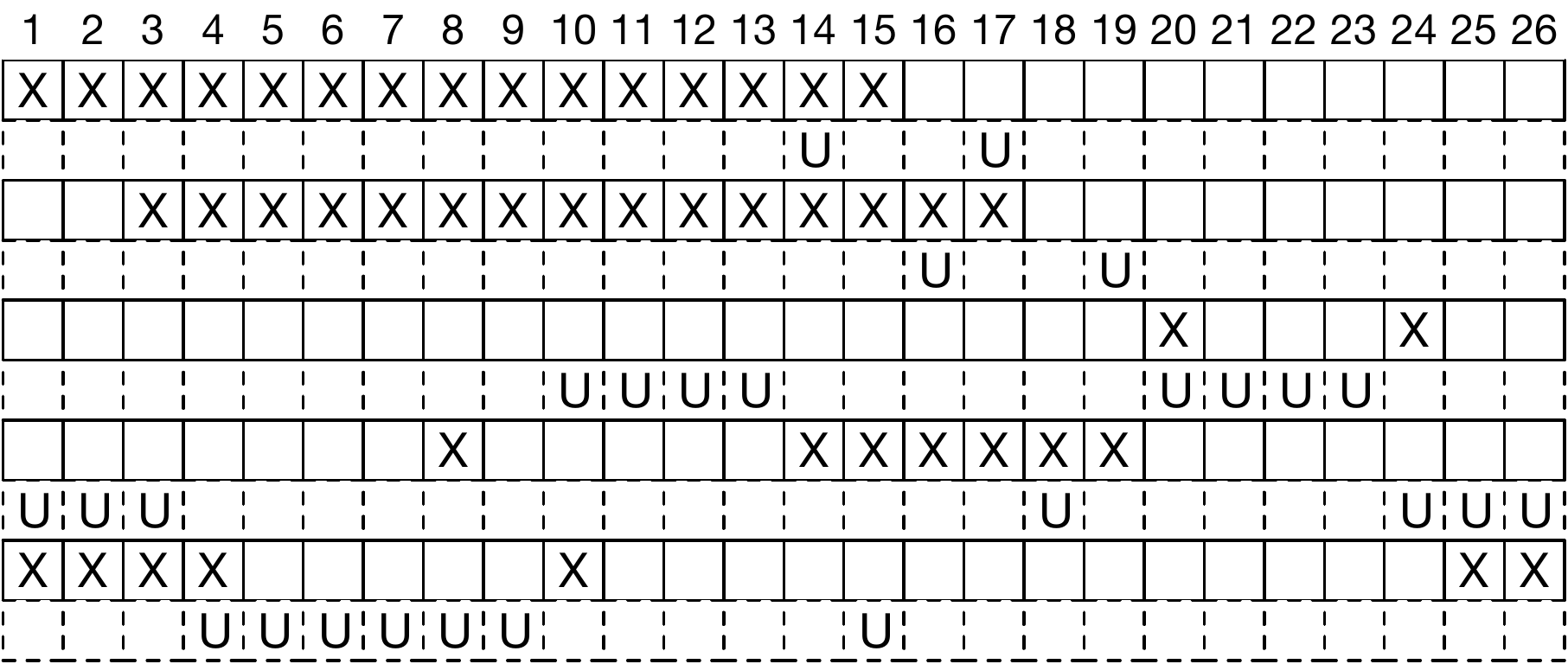}
    \caption{Achievable scheme for $m=26, s=16$.}
    \label{fig:ex:m=26,s=16}
\end{figure}

% \begin{rem}
%     \label{rem:3s/2<m<2s}
%     % In this case we use two types of schemes to satisfy the users.
%     % One important step of designing the codewords in this case is to check the security constraint.
%     In this case it takes an extra \dt{compared to? where do we see it?} step to check the security constraint since two types of schemes have been mixed.
%     {Such a step serves as a main part of the proof the proposed scheme as has been described in Appendix~\ref{sec:3s/2<m<2s}. }
% \end{rem}

We thus %proved the number of transmissions of the 
proposed schemes that are to within a constant gap from the converse bound in~\eqref{eq:lower_bound}. 
%Therefore the scheme is linearly optimal within a constant gap.
We summarize all cases and the corresponding bounds in Table~\ref{table:achievable schemes}.

\begin{table}[ht!]
\centering
\caption{All cases and the corresponding achievable schemes. 
$m_{|a}$ represents $m \bmod a$, where $a$ is an integer; 
``infea.'' is the abbreviation of ``infeasible''.
}
\begin{tabular}{||c c c c||} 
 \hline
Condition & Subcase & Converse & $\ell$  \\ [0.5ex] 
 \hline\hline
 $\frac{m}{m-s} \in \mathbb{Z}$ & {all} & $m/s$ & $m/s$ \\
 \hline
 \multirow{7}{5 em}{{$\frac{m}{m-s} \notin \mathbb{Z}$, \\$m-4 \geq s\geq 5$, \\$m > 2s$}} 
 & $m_{|2s} =0$ & 
 \multirow{6}{1 em}{{$\frac{3m}{2s}$}} 
 & $\frac{3m}{2s}$ \rule[-5pt]{0pt}{5pt} \\ 
 & $m_{|2s} =1$ &  & $\frac{3m}{2s}+\frac{4s-3}{2s}$ \rule[-5pt]{0pt}{5pt}\\ 
 & $m_{|2s} =2$ &  & $\frac{3m}{2s}+\frac{s-3}{s}$ \rule[-5pt]{0pt}{5pt}\\ 
 & $m_{|2s} =3$ &  & $\frac{3m}{2s}+\frac{6s-3}{2s}$  \rule[-5pt]{0pt}{5pt}\\ 
 & $m_{|2s} \in [4:s]$ &  & $\frac{3m}{2s} + \frac{4s-3m_{|2s}}{2s}$ \rule[-5pt]{0pt}{5pt}\\ 
 & $m_{|2s} \in [s+1:2s-2] $ &  & $\frac{3m}{2s} + \frac{6s-3m_{|2s}}{2s}$ \rule[-5pt]{0pt}{5pt}\\ 
  & $m_{|2s} =2s-1$ &  & $\frac{3m}{2s}+\frac{2s+3}{2s}$  \rule[-5pt]{0pt}{5pt}\\ 
  \hline
 \multirow{2}{5 em}{$\frac{m}{m-s} \notin \mathbb{Z}$,\\ $m-4\geq s\geq 5$,\\ $m< 2s$}  
 & $s<m\leq 3s/2$ \rule[-10pt]{0pt}{20pt} &
 \multirow{2}{1 em}{$2$}  
 & $\leq 4$ \\
 & $3s/2<m<2s$ \rule[-10pt]{0pt}{20pt} &  & $\leq 9$ \\
 % & $m=s+5$, even $s$ &  & 4 \\
 % & $m=s+4$ &  & 4 \\
 % & $m=s+3$, even $s$ &  & 4 \\ 
 %  & $m=s+3$, odd $s$ &  & 3 \\ 
  \hline
 % \multirow{6}{5em}{$\frac{m}{m-s} \notin \mathbb{Z}$,\\ $s < 5$}  
 % & $s=1$ & infeasible & $\infty$ \\
 % & $s=2$ & infeasible & $\infty$ \\
 % & $s=3$ and even $m$ & scheme 6 & $\frac{m}{2}=\frac{3m}{2s}$ \\
 % & $s=3$ and odd $m$ & infeasible & $\infty$ \\
 % & $s=4$ and even $m$ & scheme 1, 4 & $\frac{3m}{2s}+\frac{m \bmod 2s}{2s}$ \\
 % % & $s=4$ and $m\bmod{2s}=2$ & scheme 1, 4 & $\floor{\frac{3m}{2s}}+1$ \\
 % % & $s=4$ and $m\bmod{2s}=4$ & scheme 1, 4 & $\floor{\frac{3m}{2s}}+2$ \\
 % % & $s=4$ and $m\bmod{2s}=6$ & scheme 1, 4 & $\floor{\frac{3m}{2s}}+3$ \\
 % & $s=4$ and  odd $m$ & infeasible & $\infty$ \\
 % \hline
 \multirow{6}{5 em}{$s\leq 4$, $\frac{m}{m-s}\notin\mathbb{Z}$} 
 & $s=1$ & infea.  & $\infty$ \rule[-5pt]{0pt}{5pt}\\ 
& ${s} =2$ & infea. & $\infty$ \rule[-5pt]{0pt}{5pt}\\ 
& ${s} =3$, even $m$ & $\frac{m}{2}$ & $\frac{m}{2}$ \rule[-5pt]{0pt}{5pt}\\ 
& ${s} =3$, odd $m$ & infea. & $\infty$ \rule[-5pt]{0pt}{5pt}\\ 
& ${s} =4$, even $m$ & $\frac{3m}{8}$ & $3\floor{\frac{m}{8}}+\frac{m_{8}}{2}$ \rule[-5pt]{0pt}{5pt}\\ 
& ${s} =4$, odd $m$ & infea. & $\infty$ \rule[-5pt]{0pt}{5pt}\\ 
 \hline
 \multirow{3}{5 em}{$s\geq m-3$, $\frac{m}{m-s}\notin\mathbb{Z}$} 
 % & $m-s=2$ & infeasible  & $\infty$ \\ 
% & ${s} =2$ & infeasible & $\infty$ \\ 
% & ${m-s} =2$ and even $m$ & $\frac{3}{2}+\frac{3}{s}$ & $3$ \\ 
& ${m-s} =2$, odd $m$ & infea. & $\infty$ \rule[-5pt]{0pt}{5pt}\\ 
& ${m-s} =3$, even $m$ & $\frac{3}{2}+\frac{9}{2s}$ & $3$ \rule[-5pt]{0pt}{5pt}\\ 
& ${m-s} =3$, odd $m$ & $\frac{3}{2}+\frac{9}{2s}$ & $4$ \rule[-5pt]{0pt}{5pt}\\ 
 [1ex] 
 \hline
\end{tabular}
\label{table:achievable schemes}
\end{table}

\section{Conclusion}
\label{sec:summarize}

In this paper %we list and prove all infeasible cases for the 
we studied the secure decentralized PICOD with circular side information at the users. 
We first proved that some settings, unknown in the literature priori to this work, are infeasible.
For all the remaining cases, we proposes achievable schemes that use the same number of transmissions as predicted by our converse bound under the constraint of linear encoding up to a constant additive gap. 
%Therefore, our results show that the linear converse bound we proposed in~\cite{centralized_secure_picod} is tight within a constant gap.
Ongoing work includes extending the setting to other type of side information structures.

\bibliographystyle{IEEEtranS}
\bibliography{refs}

% Generated by IEEEtranS.bst, version: 1.14 (2015/08/26)
\begin{thebibliography}{1}
\providecommand{\url}[1]{#1}
\csname url@samestyle\endcsname
\providecommand{\newblock}{\relax}
\providecommand{\bibinfo}[2]{#2}
\providecommand{\BIBentrySTDinterwordspacing}{\spaceskip=0pt\relax}
\providecommand{\BIBentryALTinterwordstretchfactor}{4}
\providecommand{\BIBentryALTinterwordspacing}{\spaceskip=\fontdimen2\font plus
\BIBentryALTinterwordstretchfactor\fontdimen3\font minus
  \fontdimen4\font\relax}
\providecommand{\BIBforeignlanguage}[2]{{%
\expandafter\ifx\csname l@#1\endcsname\relax
\typeout{** WARNING: IEEEtranS.bst: No hyphenation pattern has been}%
\typeout{** loaded for the language `#1'. Using the pattern for}%
\typeout{** the default language instead.}%
\else
\language=\csname l@#1\endcsname
\fi
#2}}
\providecommand{\BIBdecl}{\relax}
\BIBdecl

\bibitem{embedded_ic}
M.~W. Alexandra~Porter, ``Embedded index coding,'' \emph{Information Theory
  Workshop}, 2019. arXiv:1904.02179.

\bibitem{index_coding_with_sideinfo}
Z.~Bar-Yossef, Y.~Birk, T.~S. Jayram, and T.~Kol, ``Index coding with side
  information,'' \emph{IEEE Trans. on Information Theory}, vol.~57, no.~3, pp.
  1479--1494, Mar 2011.

\bibitem{BrahmaFragouli-IT1115-7254174}
S.~Brahma and C.~Fragouli, ``Pliable index coding,'' \emph{IEEE Trans. on
  Information Theory}, vol.~61, no.~11, pp. 6192--6203, Nov 2015.

\bibitem{decentralized_picod}
T.~Liu and D.~Tuninetti, ``Decentralized pliable index coding,'' \emph{Proc.
  Int. Symp. Inf. Theory}, 2019.

\bibitem{private_picod}
------, ``Private pliable index coding,'' \emph{arXiv:1904.04468}, 2019.

\bibitem{centralized_secure_picod}
------, ``Secure decentralized pliable index coding,'' \emph{Proc. Int. Symp.
  Inf. Theory}, 2020.

\bibitem{private_ic}
V.~Narayanan, J.~Ravi, V.~K. Mishra, B.~K. Dey, N.~Karamchandani, and V.~M.
  Prabhakaran, ``Private index coding,'' \emph{Proc. Int. Symp. Inf. Theory},
  2018.

\bibitem{secure_picod_achievability}
S.~Sasi and B.~S. Rajan, ``On pliable index coding,'' \emph{arXiv:1901.05809},
  2019.

\bibitem{SongFragouli-ISIT2016sub-7176784}
L.~Song and C.~Fragouli, ``A deterministic algorithm for pliable index
  coding,'' in \emph{Proc. Int. Symp. Inf. Theory}, July 2016.

\end{thebibliography}

\clearpage
\appendices

% \section{No more than two transmissions}
% \label{sec:1-factor}
% When $\frac{m}{m-s}\in\mathbb{Z}$, the achievable scheme in~\cite{decentralized_picod}, which is for the decentralized PICOD without security constraint, also satisfies the security constraint. 
% This is because among all messages involved in the encoding, for each user, there is one and only one message that is not in the side information set of the user. 
% Therefore, no user is able to decode any information from the messages that are not its desired message and the defined security constraint is thus satisfied.
% The proposed achievable scheme is information theoretically optimal.
% % The number of transmissions is $\ell^*_{it}=\frac{p}{p-1}$ where $p$ is the size of the 1-factor.
% % Since there exists only one 1-factor because the NTH is regular in this case, we conclude that the optimal number of transmissions is $\ell^*_{it}=\frac{m/(m-s)}{m/(m-s)-1}=\frac{m}{s}$.
% For this case, the number of transmissions is no greater than 2. 

\section{Case $m>2s$, $ m-4 \geq s \geq 5$ }
\label{sec: m>2s}
We show that for all cases in this region, there exists a scheme that achieves the converse bound within a constant gap. 
In this region our converse bound is $3m/2s$. 
In order to achieve the optimality, we need to have the achievable scheme satisfy roughly $2s/3$ users in each transmission. 
We therefore propose  the following scheme.

\subsection{Achievable schemes}

\subsubsection{Scheme 1 satisfies $2s$ users}
\label{ssub:scheme 1}
One round of scheme 1 consists of 3 transmissions: 
\begin{enumerate}
    \item $w_{i}+w_{i+1}$,
    \item $w_{i+3}+w_{i+s}$,
    \item $w_{i+s+1}+w_{i+s+2}$. 
\end{enumerate}
  
\begin{figure}[ht]
    \centering
    \includegraphics[width=0.8\columnwidth]{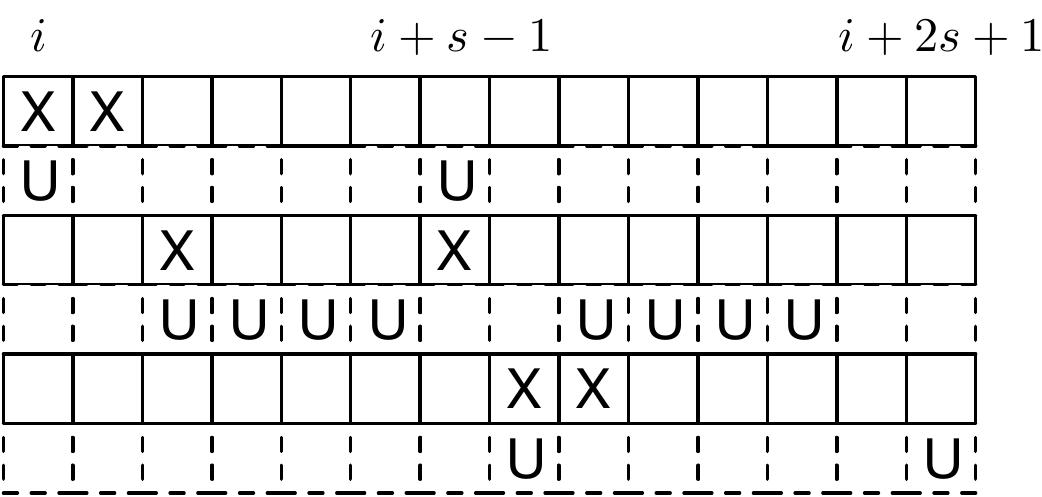}
    \caption{Scheme 1. 
    % The \textbf{X}s indicate the messages in the linear combinations. The \textbf{U}s are the users satisfied by the transmissions. 
    % The notations are used for all the figures for the following schemes. 
    }
    \label{fig:scheme1}
\end{figure}

Here $i$ is the index of the first user among the $2s$ users that are satisfied by the 3 transmissions.
% The rounds are separated by $2s$, that is, for the next round with index $j$, we have $j=i+2s$.
The $2s$ satisfied users are $u_i, U_{[i+2:i+2s-1], u_{i+2s+1}}$.
The transmissions and the satisfied users are shown in Fig.~\ref{fig:scheme1}.
As we can see the satisfied users can be seen in three groups.
The two groups on the left and right contain one user each, while
the middle group contains $2s-2$ users.
The scheme can be used repeatedly. 
The number of users that are satisfied is thus $2ks$ where $k\in\mathbb{Z}^+$.

Scheme 1 satisfies $2s$ users using 3 transmissions. 
The messages that are involved in one transmission of scheme 1 do not appear in the other transmissions of scheme 1. 
Therefore, all transmissions in all rounds of scheme 1 satisfy the security constraint.

Note that on average each transmission satisfies $2s/3$ users in this scheme.
This scheme is therefore optimal under the linear encoding constraint. 
If $\frac{m}{2s}\in\mathbb{Z}$ this is the linear optimal scheme that achieves $\ell^*=\frac{3m}{2s}$.

The schemes in the rest of this section are proposed to address the cases where $\frac{m}{2s}\notin\mathbb{Z}$ and $m>2s$.

\subsubsection{Scheme 2 for $2s+2$ users}
\label{ssub:scheme 2}
Scheme 2 satisfies $2s+2$ users using 4 transmissions. 
Let $i$ be the index of the first users that are satisfied. 
The 4 transmissions are
\begin{enumerate}
    \item $w_{i-(\ceil{\frac{1-s}{2}}-1)}+\dots+w_{i+1}$,
    \item $w_{i+2}+w_{i+\ceil{\frac{s+1}{2}}-1}$,
    \item $w_{i+\ceil{\frac{s+1}{2}}}+w_{i+s+2}$,
    \item $w_{i+s+3}+\dots+w_{i+s+\ceil{\frac{s+1}{2}}}$.
\end{enumerate}
  
\begin{figure}[ht]
    \centering
    \includegraphics[width=0.85\columnwidth]{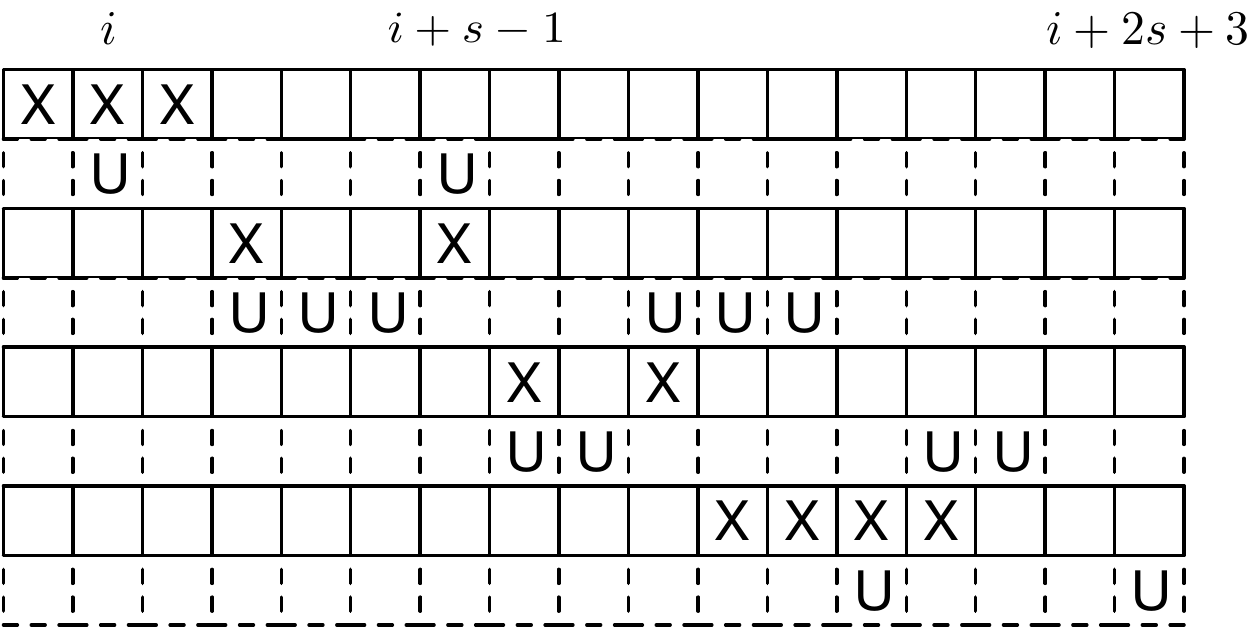}
    \caption{Scheme 2.}
    \label{fig:scheme2}
\end{figure}

% The rounds are separated by $2s$, that is, for the next round with index $j$, we have $j=i+2s$.

The satisfied users are $u_i, U_{[i+2:i+2s+1], u_{i+2s+3}}$.
The transmissions and the satisfied users are shown in Fig.~\ref{fig:scheme2}.
% On average one transmission satisfies $\frac{s+1}{2}$ users.

\subsubsection{Scheme 3 for $s+2$ users}
\label{ssub:scheme 3}
Scheme 3 satisfies $s+2$ users using 3 transmissions.
Let $i$ be the index of the first user satisfied by this scheme. 
\begin{itemize}
    \item When $s$ is even. The 3 transmissions are
    \begin{enumerate}
        \item $w_{i-s+3}+w_{i+2-s/2}+w_{i+2-s/2+1}+\dots+w_i+w_{i+1}$,
        \item $w_{i+1}+w_{i+2}+w_{i+3}$,
        \item $w_{i+3}+w_{i+4}+\dots+w_{i+s/2+2}+w_{i+s+1}$.
    \end{enumerate}
    \begin{figure}[ht]
        \centering
        \includegraphics[width=0.7\columnwidth]{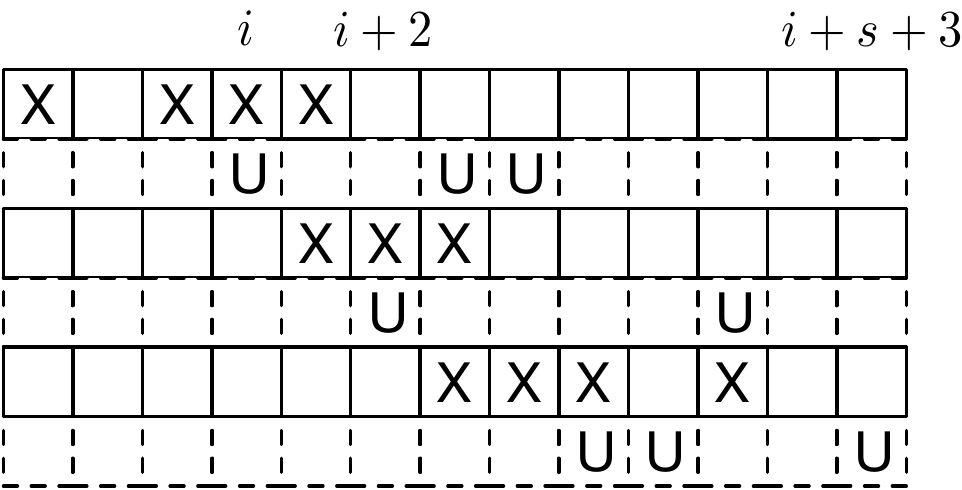}
        \caption{Scheme 3 for even $s$.}
        \label{fig:scheme3_even_s}
    \end{figure}
    \item When $s$ is odd. The 3 transmissions are 
     \begin{enumerate}
        \item $w_{i-s+4}+w_{i+2-(s-1)/2}+w_{i+2-(s-1)/2+1}+\dots+w_i+w_{i+1}$,
        \item $w_{i+1}+w_{i+2}+w_{i+4}$,
        \item $w_{i+3}+w_{i+4}+\dots+w_{i+(s+1)/2+2}+w_{i+s+1}$.
    \end{enumerate}
    \begin{figure}[ht]
        \centering
        \includegraphics[width=0.8\columnwidth]{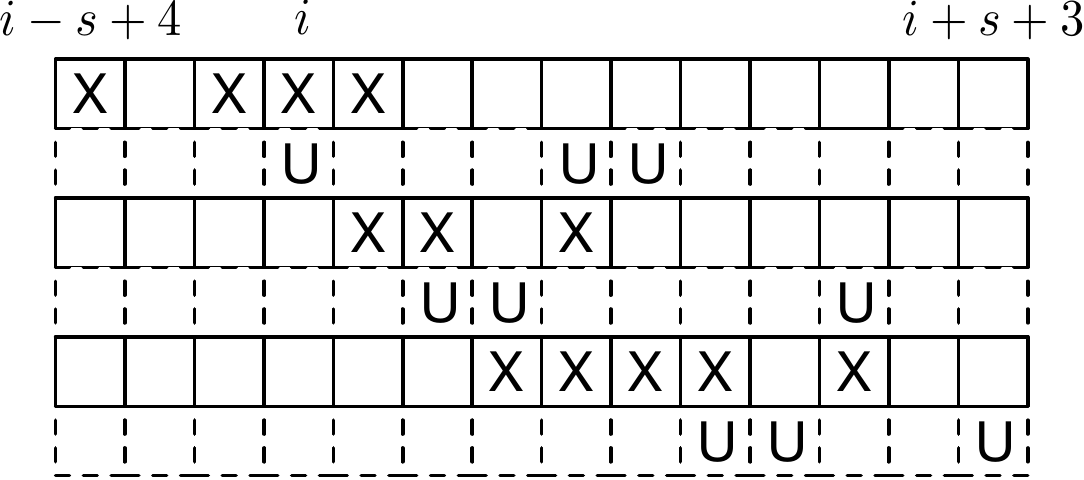}
        \caption{Scheme 3 for odd $s$.}
        \label{fig:scheme3_odd_s}
    \end{figure}
\end{itemize}

\subsubsection{Scheme 4 for $n^\prime+2\in \{2\}\cup [4:2s-2]$ users}
\label{ssub:scheme 4}
Here $n^\prime$ is the number of satisfied users in the middle group. 
\begin{itemize}
    \item $n^\prime =0$: 2 users are satisfied by one transmission:
    \begin{enumerate}
        \item $w_{i-s+3}+\dots+w_i+w_{i+1}$
    \end{enumerate}
    \begin{figure}[ht]
        \centering
        \includegraphics[width=0.35\columnwidth]{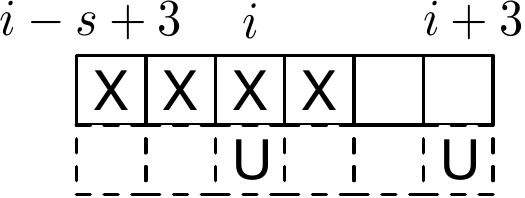}
        \caption{Scheme 4, $n^\prime=0$}
        \label{fig:scheme4_n=0}
    \end{figure}
    \item $n^\prime\in[2:s-2]$: $n^\prime+2$ users are satisfied by two transmissions:
    \begin{enumerate}
        \item $w_{i+\ceil{n^\prime/2}-s}+w_{i+n^\prime-s+2}+w_{i+n^\prime-s+3}+\dots+w_{i}+w_{i+1}$.
        \item $w_{i+n^\prime-s+3}+w_{i+n^\prime-s+4}+\dots+w_{i+2}+w_{i+\ceil{n^\prime/2}+2}$
        % +w_{i+n^\prime-s+2}+w_{i+n^\prime-s+3}+\dots+w_{i}+w_{i+1}$.
    \end{enumerate}
    \begin{figure}[ht]
        \centering
        \includegraphics[width=0.7\columnwidth]{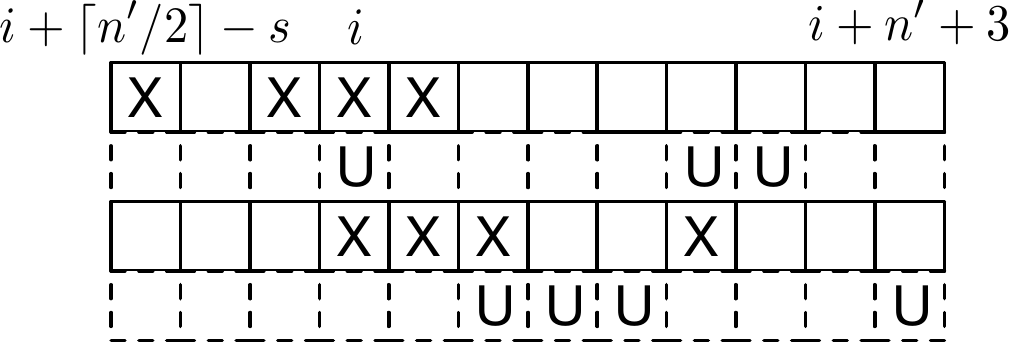}
        \caption{Scheme 4, $n^\prime\in [2:s-2]$}
        \label{fig:scheme4_n[2:n-2]}
    \end{figure}
    \item $n^\prime =s-1$: s+1 users are satisfied by three transmissions:
    \begin{enumerate}
        \item $w_{i-s+3}+w_{i+\floor{n^\prime/2}-s+2}+w_{i+\floor{n^\prime/2}-s+3}+\dots+w_i+w_{i+1}$.
        \item $w_i+w_{i+1}+w_{i+2}+w_{i+3}$.
        \item $w_{i+2}+w_{i+3}+\dots+w_{i+1+\floor{n^\prime/2}+1}+w_{i+1+n^\prime}$.
        % w_{i-s+3}+w_{i+\floor{n^\prime/2}-s+2}+w_{i+\floor{n^\prime/2}-s+3}+\dots+w_i+w_{i+1}$.
    \end{enumerate}
    \begin{figure}[ht]
        \centering
        \includegraphics[width=0.8\columnwidth]{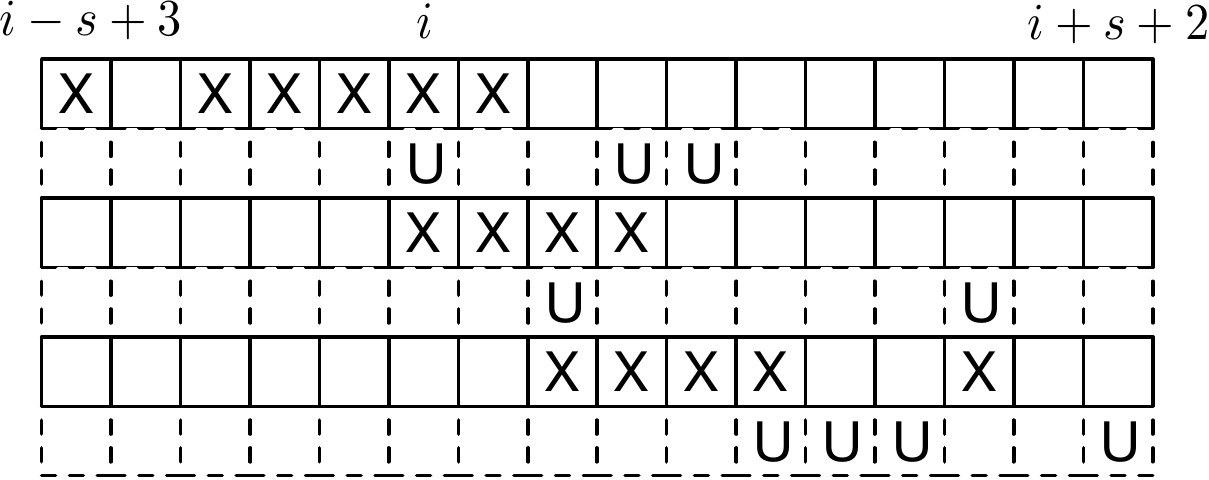}
        \caption{Scheme 4, $n^\prime=s-1$}
        \label{fig:scheme4_n=s-1}
    \end{figure}
    \item $n^\prime =s$: s+2 users are satisfied by three transmissions:
    \begin{enumerate}
        \item $w_{i-s+3}+w_{i+\floor{n^\prime/2}-s+2}+w_{i+\floor{n^\prime/2}-s+3}+\dots+w_i+w_{i+1}$.
        \item $w_{i+1}+w_{i+2}+w_{i+3}$.
        \item $w_{i+3}+\dots+w_{i+1+\floor{n^\prime/2}+1}+w_{i+1+n^\prime}$.
        % w_{i-s+3}+w_{i+\floor{n^\prime/2}-s+2}+w_{i+\floor{n^\prime/2}-s+3}+\dots+w_i+w_{i+1}$.
    \end{enumerate}
    \begin{figure}[ht]
        \centering
        \includegraphics[width=0.8\columnwidth]{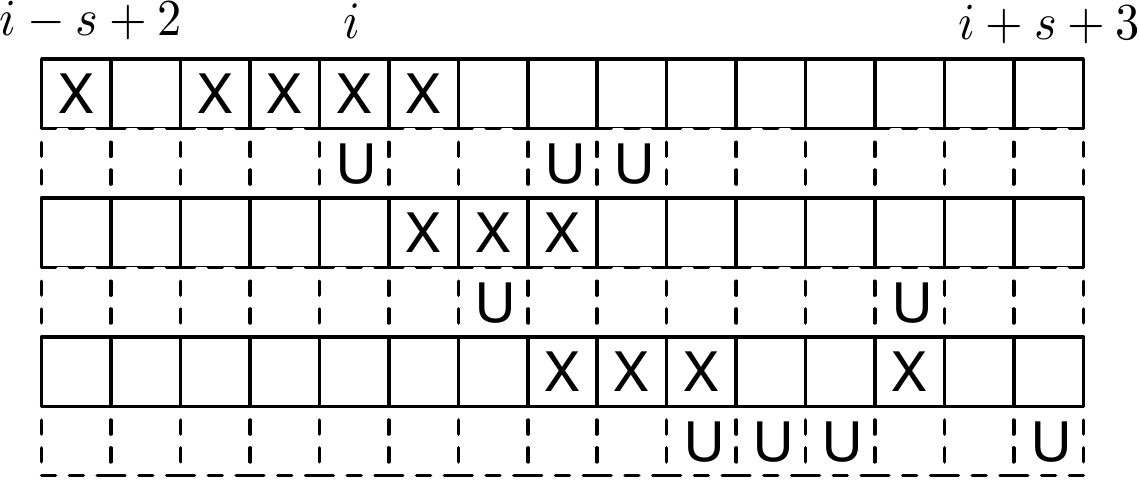}
        \caption{Scheme 4, $n^\prime=s$}
        \label{fig:scheme4_n=s}
    \end{figure}
    \item $n^\prime \in [s+1: 2s-4]$: $n^\prime+2$ users are satisfied by three transmissions:
    \begin{enumerate}
        \item $w_{i+1+n^\prime-s-s} + w_{i+\floor{n^\prime/2}-s+2}+w_{i+\floor{n^\prime/2}-s+3}+\dots+w_i+w_{i+1}$.
        \item $w_{i+2}+w_{i+1+n^\prime-s}$.
        \item $w_{i+1+n^\prime-s+1}+w_{i+1+n^\prime-s+2}+\dots+w_{i+1+\floor{n^\prime/2}+1}+w_{i+2+s}$.
        % w_{i-s+3}+w_{i+\floor{n^\prime/2}-s+2}+w_{i+\floor{n^\prime/2}-s+3}+\dots+w_i+w_{i+1}$.
    \end{enumerate}
    \begin{figure}[ht]
        \centering
        \includegraphics[width=0.7\columnwidth]{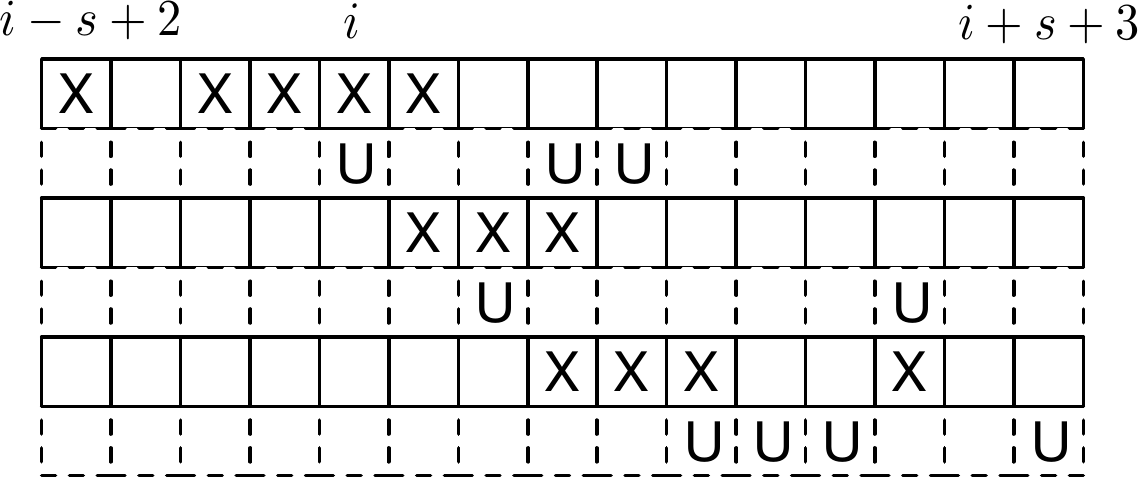}
        \caption{Scheme 4, $n^\prime\in [s+1:2s-4]$}
        \label{fig:scheme4_n[s+1:2s-4]}
    \end{figure}
\end{itemize}

\subsection{Decodability}
\label{sub:decodability}
We the proposed schemes can satisfy all $m>2s$ users in the system.

% \subsubsection{$m/(m-s)\in \mathbb{Z}$}
% In this case we have 1-factor in the NTH. 
% That is, the optimal decentralized scheme for the problem setting without security constraint is in fact secure. Therefore the scheme is information theoretical optimal.
% We have $\ell^*_{it}= m/s$.

% \subsubsection{$m\geq 2s, s\geq 5$}
To achieve the converse bound $\ell \geq 3m/2s$, we aim to satisfy $2s/3$ users in one transmission. 
Therefore, the intuition is to use scheme 1 as much as possible. 
Then we use schemes that involve a constant number of transmissions to satisfy the remaining users. 

Note that scheme 1 satisfies $2s$ users in each round, so the number of remaining users is less than $2s$ after using scheme 1. 
Scheme 4  can satisfy users of size $2$ or $[4:2s-2]$. 
By applying scheme 4 after scheme 1, we can deal with the cases where $m \bmod{2s} \in \{0,2,4,5,\dots,2s-2\}$.
For the case $m \bmod{2s} = 2s-1$, we use possibly scheme 1, then scheme 2 and scheme 4.
For the cases $m \bmod{2s} \in \{1,3\}$, we use possibly scheme 1, then scheme 3 and scheme 4.

The required numbers of transmissions are addressed as the follows.
\begin{itemize}
    \item $m\bmod{2s} \neq 1,3,2s-1$. In this case we use scheme 1 $k$ times to satisfy $2sk$ users.
    The remaining users are satisfied by scheme 4.
    The total number of transmissions used is 
    \begin{align}
        % \ell = \begin{cases}
        % 3\floor{\frac{m}{2s}}+1=\frac{3m}{2s}+\frac{s-3}{s} & m \bmod{2s}=2 \\
        % 3\floor{\frac{m}{2s}}+2=\frac{3m}{2s}+\frac{4s-3(m\bmod 2s)}{2s} & m \bmod{2s}\in [4:s] \\
        % 3\floor{\frac{m}{2s}}+3=\frac{3m}{2s}+\frac{6s-3(m\bmod 2s)}{2s} & m \bmod{2s} \in [s+1:2s-2] \\
        % \end{cases}
        \ell = \begin{cases}
        3\floor{\frac{m}{2s}}+1 & m \bmod{2s}=2 \\
        3\floor{\frac{m}{2s}}+2 & m \bmod{2s}\in [4:s] \\
        3\floor{\frac{m}{2s}}+3 & m \bmod{2s} \in [s+1:2s-2] \\
        \end{cases}
    \end{align}
    \item $m \bmod{2s} = 2s-1$. In this case $m\geq 4s-1$. We first use scheme 1 until the number of remaining users becomes $4s-1$. We then use scheme 2 to satisfy $2s+2$ users. The number of remaining users after applying scheme 2 is $2s-3$. We then use scheme 4 to satisfy these $2s-3$ users. 
    The total number of transmissions  is
    \begin{align}
    \ell = 3\frac{m+1}{2s}+1=\frac{3m}{2s}+\frac{2s+3}{2s}.
    \end{align}
    \item $m \bmod{2s} = 1$. In this case $m\geq 2s+1$. We first use scheme 1 until the number of remaining users becomes $2s+1$. We then use scheme 3 to satisfy $s+2$ users. The number of remaining users after applying scheme 3 is $s-1$. We then use scheme 4 to satisfy these $s-1$ users. 
    The total number of transmissions is
    \begin{align}
    \ell = 3\frac{m-2s-1}{2s}+3+2=\frac{3m}{2s}+\frac{4s-3}{2s}.
    \end{align}
    \item $m \bmod{2s} = 3$. In this case $m\geq 2s+3$. We first use scheme 1 until the number of remaining users becomes $2s+3$. We then use scheme 3 to satisfy $s+2$ users. The number of remaining users after applying scheme 2 is $s+1$. We then use scheme 4 to satisfy these $s+1$ users. 
    The total number of transmissions  is
    \begin{align}
    \ell = 3\frac{m-2s-1}{2s}+3+3=\frac{3m}{2s}+\frac{6s-3}{2s}.
    \end{align}
\end{itemize}

\subsection{Security constraint}
\label{sub:security}
% In section~\ref{sub:decodability} we settled the decodability of the problem, i.e., all users can decode one message that is not in their side information sets.
In this part we show that the proposed schemes are secure. 
% That is, each user can not decode any message other than its desired message. 
%  each transmission is a linear combination of the messages that are in the side information set of one user in the system. 
By definition, the scheme is secure if all users can not decode any single message after decoding the desired messages. 
Since the encoding function is linear, the security constraint is equivalent to that there does not exist any standard basis in the row span of the submatrix of the generation matrix which is reduced by the columns which correspond to the interfering messages of a user.
 % by taking the columns corresponding to the interfering messages, there does not exist any standard basis. 

% \subsubsection{Security within the scheme}
% \label{ssub:secure_within_scheme}

% We show that the schemes we proposed are secure in the sense that the users will not decode more than one messages by one of the schemes. 
We details the arguments for each scheme. 
\begin{enumerate}
    \item Scheme 1 satisfies $2s$ users using 3 transmissions.  
    The messages in different transmissions are all different. 
    Therefore, the submatrix obtained from the columns of interfering messages of a user has all rows vectors orthogonal. 
    None of the row vectors are standard basis. 
    Therefore, the row span of the submatrix does not include any standard basis.
    The scheme 1 is thus secure.

    Note that it is also secure to repeat scheme 1 repeatedly since the row span of the submatrix obtained for one user does not contain any standard basis as well. 
    This shows that the case $m \bmod 2s = 0$ can be satisfied securely using scheme 1.

    When followed by other schemes, the transmissions of scheme 1 also do not share any common messages to the transmissions of the other schemes. 
    Therefore, the security holds for the case $m \bmod 2s = 2$ and $m \bmod 2s \in [4:2s-2]$.

    \item Scheme 2 satisfies $2s+2$ users using 4 transmissions.
    Note that all 4 transmissions do not have common messages involved. 
    By the same argument for scheme 1, the $2s+2$ users that are satisfied by scheme 2 will not decode more than one message by the transmissions in scheme 2.

    When followed by scheme 4, there are exactly two transmissions which shared an exactly one message. 
    Since the transmissions evolve strictly more than 2 messages, the interfering subspace does not contain any standard basis and the security constraint still holds.

    \item Scheme 3 satisfies $s+2$ users using 3 transmissions.
    The 3 transmissions have at most one message in common pairwise. 
    The 3 rows in the submatrix by the interfering messages contains at least 2 non-zero elements. Therefore, the linear combination of the rows can not be the standard basis. 
    Therefore, the row span of the submatrix does not contain any standard basis. The $s+2$ users that are satisfied by scheme 3 will not decode more than one message by the transmissions in scheme 3.

    Again, when followed by scheme 4, one can check that the interfering subspace does not contain any standard basis.
    The user can not decode any message other than its desired message.
    The security constraint is satisfied.

    \item Scheme 4 uses different number of transmissions for different subcases
    \begin{itemize}
        \item  When $n^\prime=0$, 2 users are satisfied by 1 transmission. It is thus secure.
        \item When $n^\prime\in [2:s-2]$,  the scheme involves 2 transmissions. There are 2 messages in common of these 2 transmissions. 
        Since each row contains at least 4 non-zero elements, any standard basis can not be a linear combination of the 2 rows. 
        \item When  $n^\prime =s-1$,  the scheme involves 3 transmissions. There are at most 2 messages in common pairwise. 
        Since each row contains at least 4 non-zero elements, any standard basis can not be a linear combination of the 3 rows. 
        \item When  $n^\prime =s$,  the scheme involves 3 transmissions. There is at most 1 message in common pairwise. 
        Since each row contains at least 3 non-zero elements, any standard basis can not be a linear combination of the 3 rows.
        \item When  $n^\prime \in [s+1:2s-4]$,  the scheme involves 3 transmissions. There is no message in common of these 3 transmissions. 
        % Since each row contains at least 3 non-zero elements, any standard basis can not be a linear combination of the 3 rows.
        It is thus secure.
    \end{itemize}
\end{enumerate}
Overall,  we conclude that no user is going to decode more than one message. 
The proposed scheme is secure.

\section{Case $s<m\leq 3s/2$, $ m-4 \geq s \geq 5$}
\label{sec:s<m<3s/2}

In this regime we consider another type of achievable scheme. 
We generate the codewords that satisfy the following two conditions:
\begin{enumerate}
    \item After subtracting the messages in the side information set, each user observes one and only one transmission that contains only one message.
    % , which is the desired message of the user.
    \item No common message between any two transmissions. That is, one message only appears in at most on transmission.
\end{enumerate}
These two conditions guarantee the decodability as well as the security since linear combinations of the transmissions do not help to decode any new message. 
Therefore, the security per transmission provides the security of the whole transmissions.

\subsection{Case $m=k_{1}(2p-2)+k_{2}(p)$}
\label{sub:m=k_{1}(2p-2)+k_{2}(p)}
For the simplicity of notation, define $p:=m-s$, which is the size of the complement of side information set.
We first provide a basic structure of the codewords, which works for the case $m=k_{1}(2m-2s-2)+k_{2}(m-s)$, where $k_{1}\geq 2, k_2\geq 0, k_1,k_2\in\mathbb{Z}$.

The scheme takes 4 transmissions:
\begin{itemize}
    \item $\sum_{i=1}^{p-1}w_i + \sum_{j=2}^{k_1+k_2-1}\sum_{i=1}^{p-1} w_{2j(2p-2)+i}$.
    \item $w_{p} + w_{2p-2} + \sum_{j=2}^{k_1-1}(w_{j(2p-2)+p}+w_{(j+1)(2p-2)})+\sum_{j=0}^{k_2-1} w_{m-jp}$.
    \item $\sum_{i=2p-1}^{3p-3}w_i$.
    \item $w_{3p-2}+w_{4p-4}$.
\end{itemize}
The transmissions are illustrated in Fig.~\ref{fig:case_m=k_{1}(2p-2)+k_{2}(p)}.
    \begin{figure}[ht]
        \centering
        \includegraphics[width=\columnwidth]{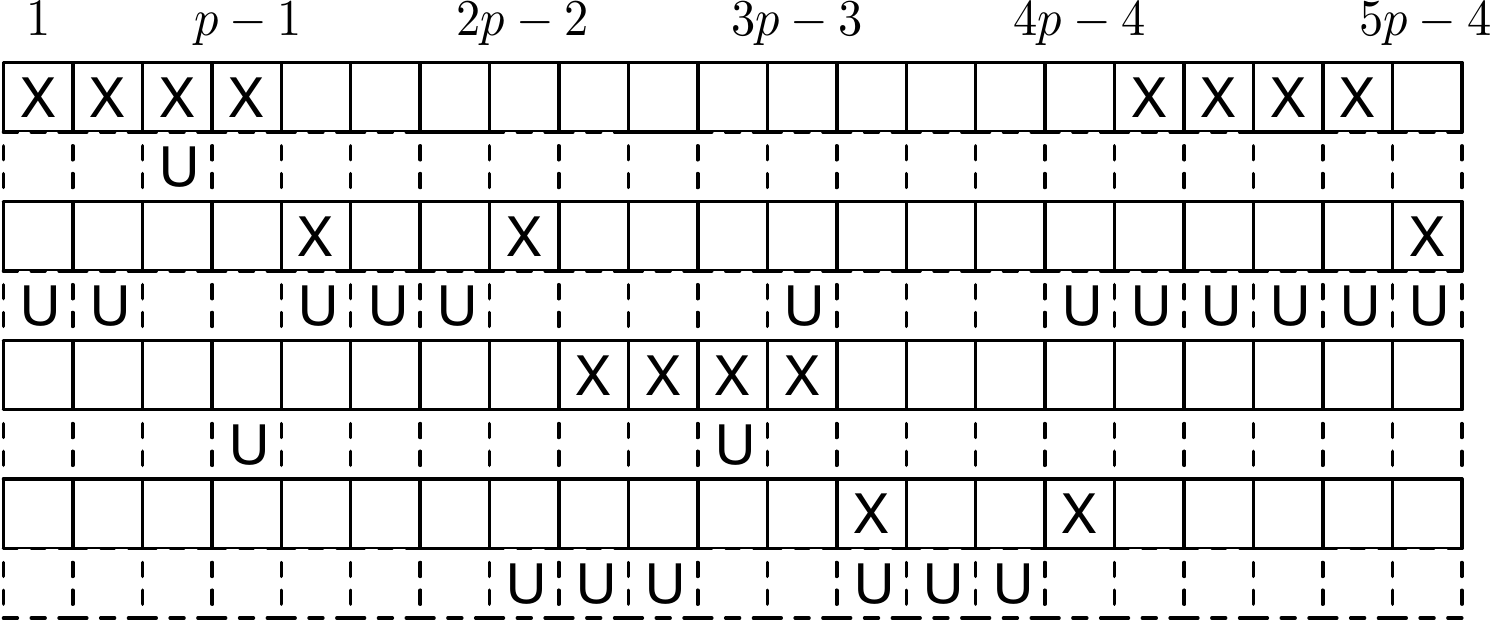}
        \caption{Scheme for case $m=k_{1}(2p-2)+k_{2}(p)$}
        \label{fig:case_m=k_{1}(2p-2)+k_{2}(p)}
    \end{figure}

By manipulating the parameters $k_1,k_2$, the scheme can satisfy all cases of $m\geq 4p$ except the following cases:
\begin{itemize}
    \item $m=5p-1, 5p-2, 7p-3$.
    \item $m=5p-3, 6p-3$.
    % \item $m=$.
\end{itemize}

\subsection{Case $m=k_2(2p-2)+q$}
\label{sub:m=k_2(2p-2)+k_2p+q}

For the proposed scheme for case $m=k_{1}(2p-2)+k_{2}(p)$ in Appendix~\ref{sub:m=k_{1}(2p-2)+k_{2}(p)},
call the part associated with $k_1$ part A, which consists of blocks of size $2p-2$; the part associated with $k_2$ part B, which consists of blocks of size $p$.
We modify the scheme such that it works for $m=k_2(2p-2)+q$, where $k_{1}\geq 1, p+2\leq q \leq 2p-2, k_1, q\in\mathbb{Z}$.
The modification is done by shrink one block in part A to size $q$. 

\begin{figure}[ht]
        \centering
        \includegraphics[width=0.8\columnwidth]{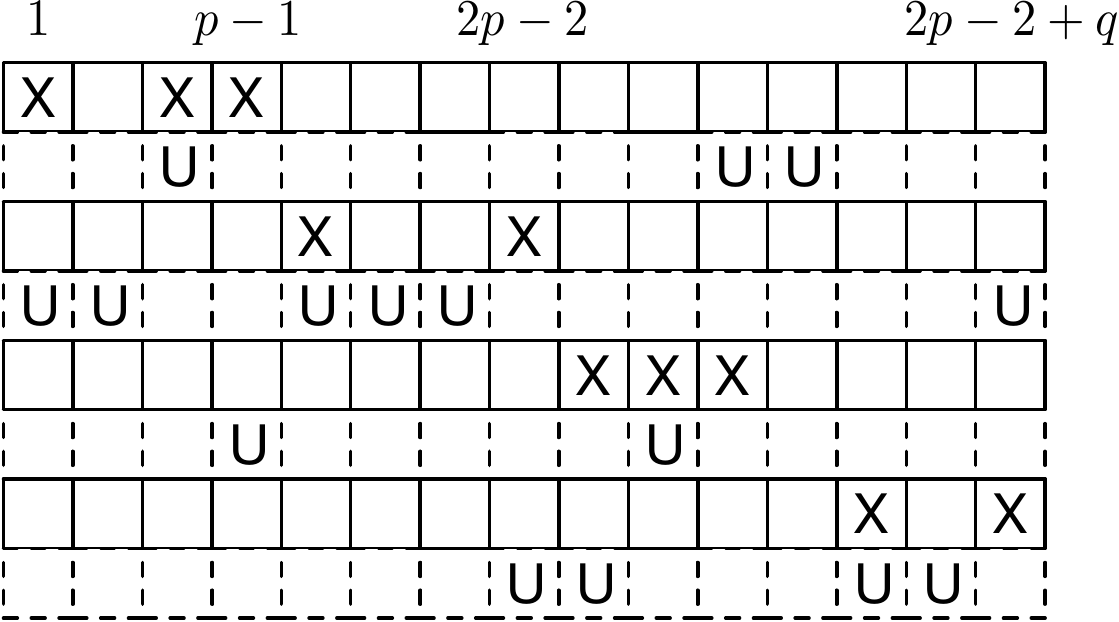}
        \caption{Scheme for case $m=k_2(2p-2)+q$}
        \label{fig:case_m=k_1(2p-2)+q}
    \end{figure}

The scheme takes 4 transmissions:
\begin{itemize}
    \item $w_{1}+w_{2p-q}+\dots+w_{p-1}+\sum_{j=1}^{k_1-1}\sum_{i=1}^{p-1}w_{j(2p-2)+i}$.
    \item $w_{p} + w_{2p-2} + \sum_{j=1}^{k_1-1}(w_{j(2p-2)+p}+w_{j(2p-2)+2p-2})$.
    \item $\sum_{i=m-q+1}^{m-p+1}w_i$.
    \item $w_{m-q+p}+w_{m}$.
\end{itemize}
The transmissions are illustrated in Fig.~\ref{fig:case_m=k_1(2p-2)+q}.

By taking $k_1=2,q=p+3$, $k_1=2,q=p+2$, $k_1=3,q=p+3$, the cases $m=5p-1, 5p-2, 7p-3$ can be solved using the this scheme. 
Moreover, by taking $k_1=1$, the case $3p \leq m \leq 4p-4$ can be solved by this scheme as well.

\subsection{Case by case study}
\label{sub:case by case}

We are now left with the case $m=5p-3, 6p-3$ and $m= 4p-3, 4p-2, 4p-1$.
Since there are only 5 cases, we address them case by case.

\subsubsection{$m=5p-3$}

\begin{figure}[ht]
        \centering
        \includegraphics[width=\columnwidth]{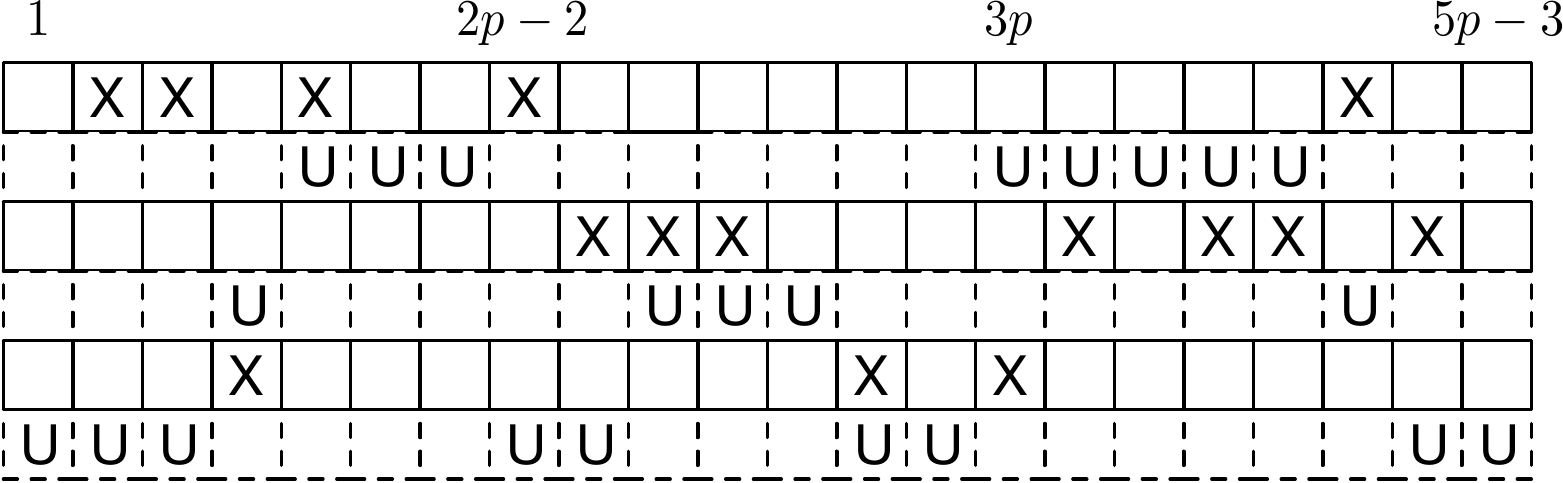}
        \caption{Scheme for case $m=5p-3$}
        \label{fig:case_m=5p-3}
    \end{figure}

The scheme takes 3 transmissions:
\begin{itemize}
    \item $\sum_{i=2}^{p-2}w_i+w_{p}+w_{2p-2}+w_{4p}$.
    \item $\sum_{i=2p-1}^{2p+1}w_i+w_{3p+1}+w_{4p-2}+w_{4p-1}+w_{5p-4}$.
    \item $w_{p-1}+w_{3p-2}+w_{3p}$.
\end{itemize}
The transmissions are illustrated in Fig.~\ref{fig:case_m=5p-3}.

\subsubsection{$m=6p-3$}

\begin{figure}[ht]
        \centering
        \includegraphics[width=\columnwidth]{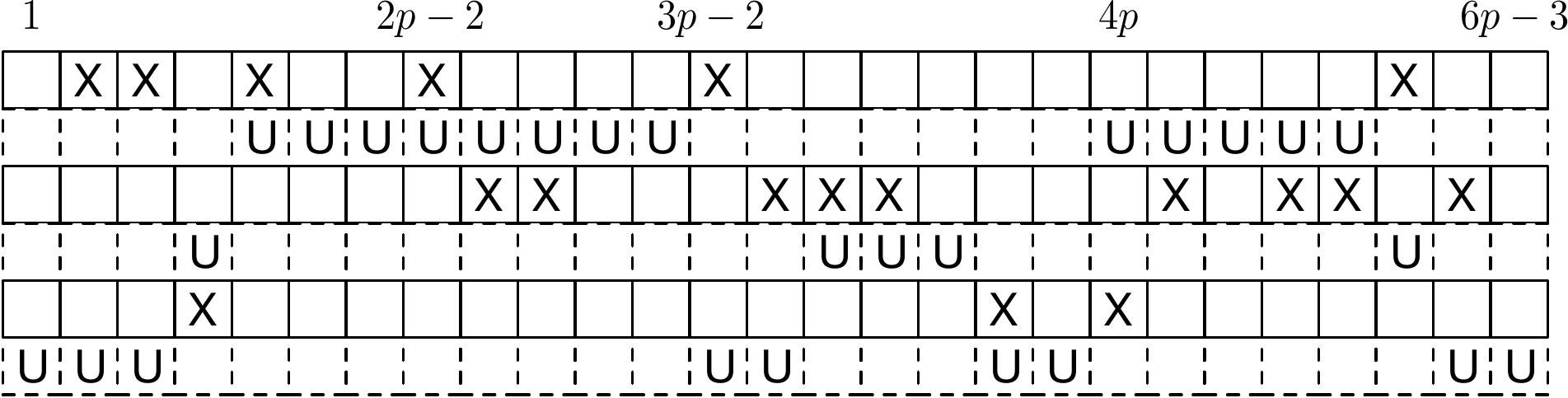}
        \caption{Scheme for case $m=6p-3$}
        \label{fig:case_m=6p-3}
    \end{figure}

The scheme can be seen as modified scheme for the case $m=5p-3$ by adding a block of size $p$.
The scheme takes 3 transmissions:
\begin{itemize}
    \item $\sum_{i=2}^{p-2}w_i+w_{p}+w_{2p-2}+w_{3p-2}+w_{5p}$.
    \item $\sum_{j=0}^{1}\sum_{i=2p-1}^{2p+1}w_{i+jp}+w_{4p+1}+w_{5p-2}+w_{5p-1}+w_{6p-4}$.
    \item $w_{p-1}+w_{4p-2}+w_{4p}$.
\end{itemize}
The transmissions are illustrated in Fig.~\ref{fig:case_m=6p-3}.

\subsubsection{$m=4p-3$}
\begin{figure}[ht]
        \centering
        \includegraphics[width=0.8\columnwidth]{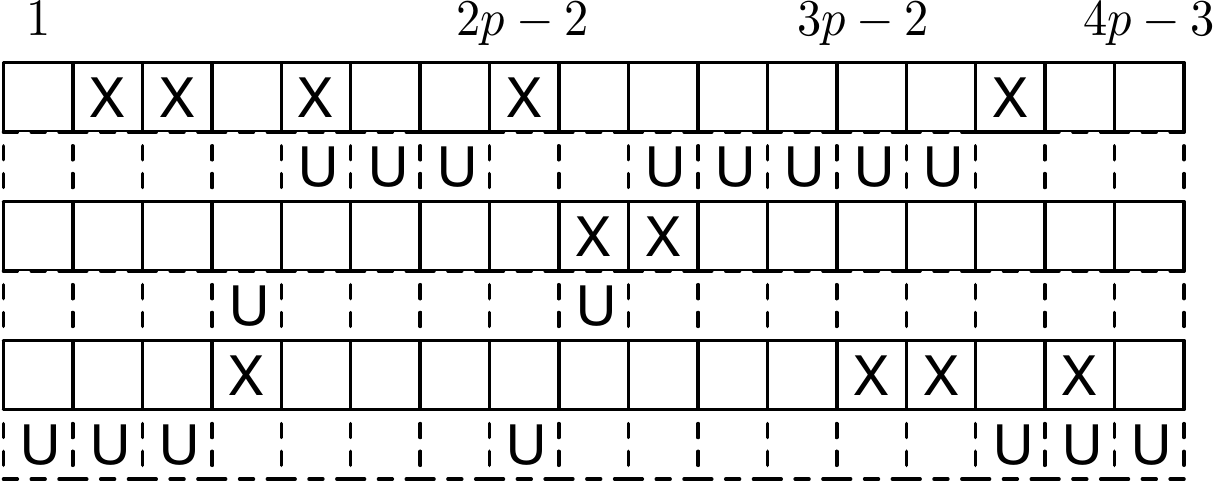}
        \caption{Scheme for case $m=4p-3$}
        \label{fig:case_m=4p-3}
    \end{figure}

The scheme takes 3 transmissions:
\begin{itemize}
    \item $w_{p-1}+w_{3p-2}+w_{3p-1}+w_{4p-4}$ or $w_{p-1}+w_{3p-2}+w_{3p-1}$ if $p=4$.
    \item $\sum_{i=2}^{p-2}w_{i}+w_{p}+w_{2p-2}+w_{3p}$.
    \item $w_{2p-1}+w_{2p}$.
\end{itemize}
The transmissions are illustrated in Fig.~\ref{fig:case_m=4p-3}.

\subsubsection{$m=4p-2$}
\begin{figure}[ht]
        \centering
        \includegraphics[width=0.8\columnwidth]{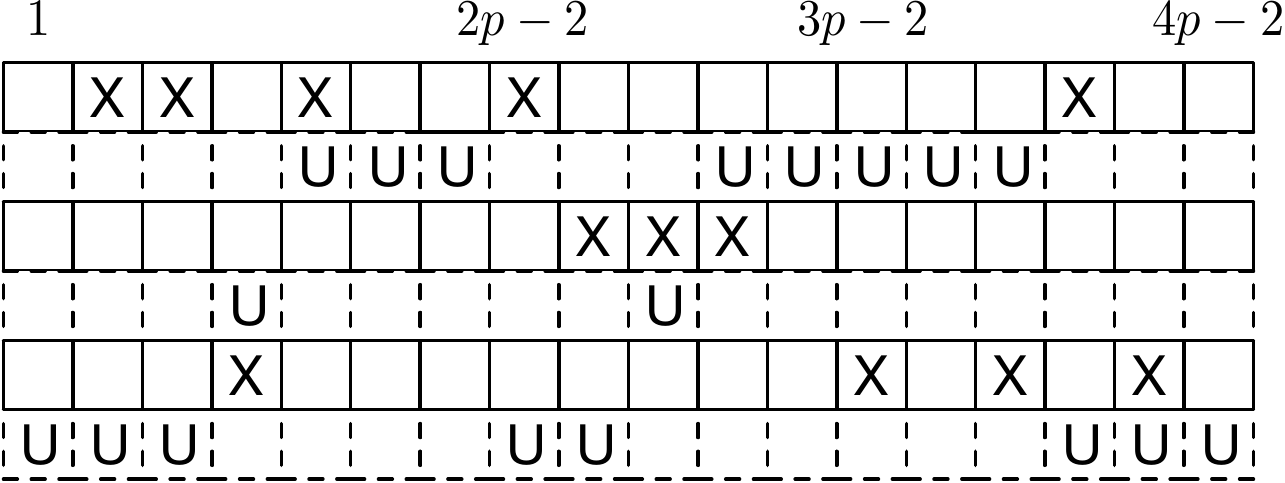}
        \caption{Scheme for case $m=4p-2$}
        \label{fig:case_m=4p-2}
    \end{figure}

The scheme takes 3 transmissions:
\begin{itemize}
    \item $w_{p-1}+w_{3p-2}+w_{3p}+w_{4p-3}$ or $w_{p-1}+w_{3p-2}+w_{3p}$ if $p=4$.
    \item $\sum_{i=2}^{p-2}w_{i}+w_{p}+w_{2p-2}+w_{3p+1}$.
    \item $w_{2p-1}+w_{2p}+w_{2p+1}$.
\end{itemize}
The transmissions are illustrated in Fig.~\ref{fig:case_m=4p-2}.

\subsubsection{$m=4p-1$}

\begin{figure}[ht]
        \centering
        \includegraphics[width=0.8\columnwidth]{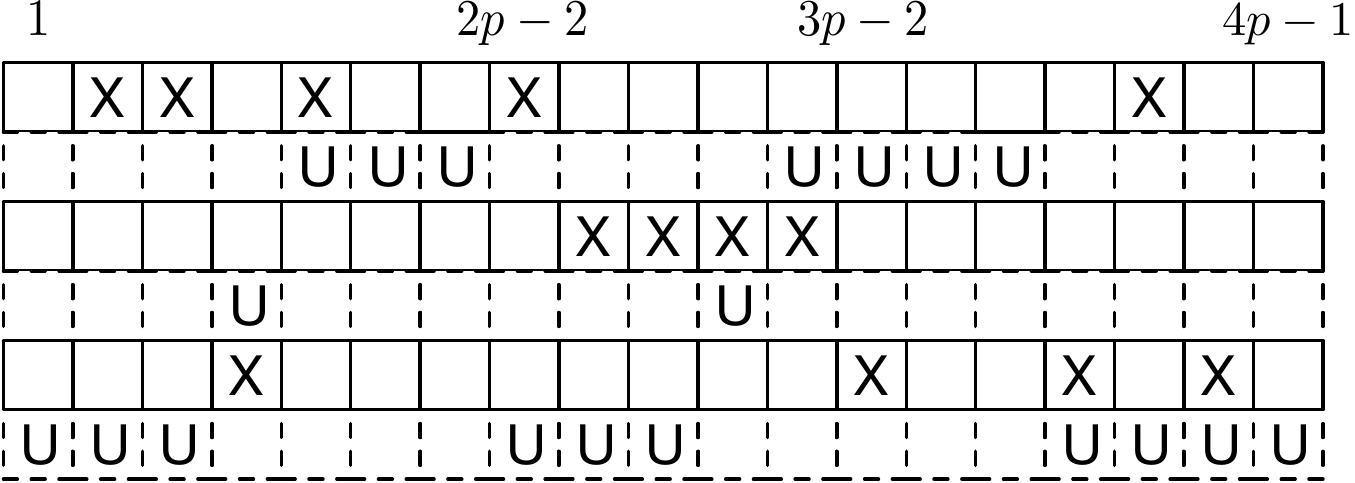}
        \caption{Scheme for case $m=4p-1$}
        \label{fig:case_m=4p-1}
    \end{figure}

The scheme takes 3 transmissions:
\begin{itemize}
    \item $w_{p-1}+w_{3p-2}+w_{3p+1}+w_{4p-2}$ or $w_{p-1}+w_{3p-2}+w_{3p+1}$ if $p=4$.
    \item $\sum_{i=2}^{p-2}w_{i}+w_{p}+w_{2p-2}+w_{3p+2}$.
    \item $w_{2p-1}+w_{2p}+w_{2p+1}+w_{2p+2}$.
\end{itemize}
The transmissions are illustrated in Fig.~\ref{fig:case_m=4p-1}.

Therefore, all the cases in the regime $s<m\leq 3s/2, 4\leq s \leq m-5$ can be satisfied securely by the proposed scheme in this section using a finite number of transmissions.
The linear converse bound can be achieved within an additive constant gap
% In this regime the converse bound $\ell\geq \frac{3m}{2s}$ is not valid anymore. 
% We aim to show that a constant number of transmissions can satisfy all users for the feasible cases.
% We use scheme 3 first to satisfy $s+2$ users. To use scheme 3 we need $m\geq s+4$. The number of remaining users after scheme 3 is $2 \leq m-s-2<s-2$. 
% To apply scheme 4, we need $m-s-2 \neq 1,3,2s-1$. That is, $m\neq s+3, s+5$.
% The total number of transmissions used is 
% \begin{align}
%     \ell = \begin{cases}
%     5, & s+6 \leq m < 2s,\\
%     4, & m = s+4.
%     \end{cases}
% \end{align}

% When $m=s+5$ and $s$ is odd, we use the scheme 6. The number of transmissions is $4$.
% When $m=s+5$ and $s$ is even, we use scheme 5. The number of transmissions needed is $4$.

% When $m=s+3$ and $s$ is odd, we use the scheme 6. The number of transmissions is $3$.
% When $m=s+3$ and $s\geq 8$ is even, we use scheme 7. the number of transmissions is $4$.
% When $m=s+3$ and even $s<8$. We have cases $(s=2,m=5)$, $(s=4,m=7)$, $(s=6,m=9)$. The first two cases are infeasible and in the last case $m/(m-s)=3\in\mathbb{Z}$.

% Therefore, for the case where $m<2s, s\geq 5$, the number of transmissions to satisfy all users are
% \begin{align}
%     \ell = \begin{cases}
%     % m/s, & m/(m-s)\in\mathbb{Z},\\
%     5, &  s+6 \leq m < 2s,\\
%     4, &  m = s+4, s+5, \\
%     4, & m=s+3, \text{even } s\geq 8 \\
%     3, & m=s+3, \text{odd } s \\
%     \text{infeasible}, & \text{otherwise}.
%     \end{cases}
% \end{align}

\section{Case $3s/2<m < 2s$, $ m-4 \geq s \geq 5$}
\label{sec:3s/2<m<2s}

In this regime we aim to satisfy $m$ users securely by a constant number of transmissions. 
The idea is to satisfy a small amount of the users at first such that the remaining users can be satisfied using the existing scheme securely. 
Specifically,  we want to have the remaining users such that the schemes for the case $m\geq 2s$ can be applied and the security constraint can be satisfied
We use the notation $p:=m-s$ which is defined in Appendix~\ref{sec:s<m<3s/2}.

\subsection{Case $p\geq 7$}
The scheme we proposed can be split into two parts.
The first part contains $a$ transmissions, where $a=2$ or $3$ and will be decided later.
The transmissions are $\sum_{i=1}^{s-1}w_i, \sum_{i=3}^{s+1}w_i$ when $a=2$ and $\sum_{i=1}^{s-1}w_i, \sum_{i=3}^{s+1}w_i, \sum_{i=5}^{s+3}w_i$ when $a=3$.
% \begin{itemize}
%     \item $\sum_{i=1}^{s-1}$.
%     \item $\sum_{i=3}^{s+1}$
%     \item $\sum_{i=5}^{s+3}$
% \end{itemize}
Therefore, the first part satisfies $2a$ users. The number of remaining users are $m-2a$. 

\begin{figure}[ht]
    \centering
    \includegraphics[width=0.95\columnwidth]{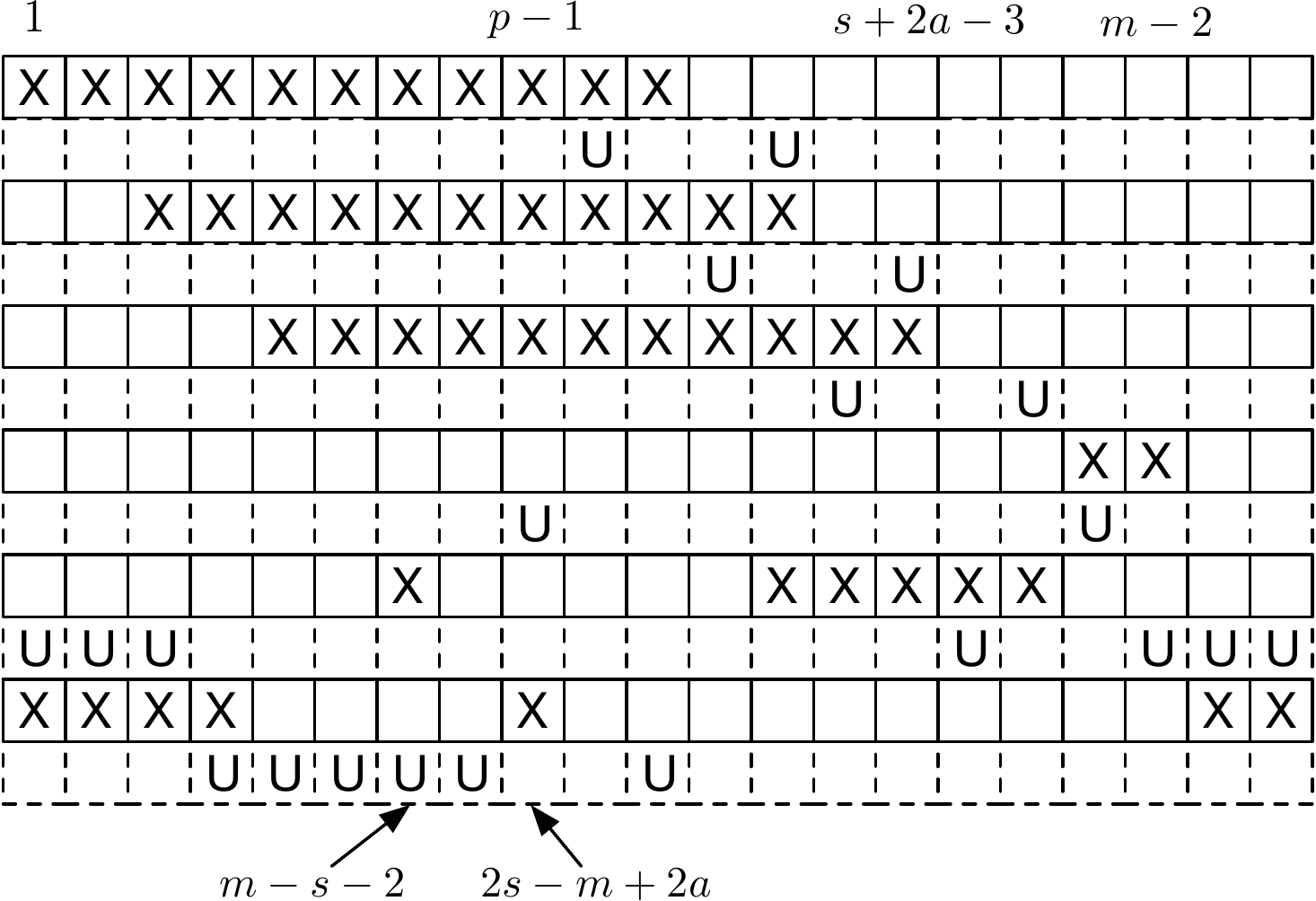}
    \caption{Scheme for the case $3s/2<m<2s$}
    \label{fig:scheme_pgeq7}
\end{figure}

In the second part, we use the scheme that is similar to scheme 4 in Appendix~\ref{ssub:scheme 4} that has been used for case $m\geq 2s$.
The second part takes 3 transmissions as the follows.
\begin{itemize}
    \item $w_{s+2a}+w_{m-2}$.
    \item $w_{m-s-2}+\sum_{i=a_{1}+\ceil{\frac{m}{2}}-1}^{s+2a-1}w_i$.
    \item $\sum_{i=1}^{s+a-\floor{\frac{m}{2}}-1}w_i+w_{2s-m+2a}+w_{m-1}+w_{m}$.
\end{itemize}
Note that the first transmission needs $m-2\geq s+2a+1$, which implies the condition $m-s=p\geq 2a+3\geq 7$.
By the 3 transmissions, the second part satisfies all the remaining $m-2a$ users. 
The schemes are illustrated in Fig.\ref{fig:scheme_pgeq7}.

We now decide the value of $a$, which will make the security constraint satisfied.

The security constraint can not be satisfied when the second part allows some users to decode more than one message.
Note that the transmissions in the second part are the scheme used in Appendix~\ref{ssub:scheme 4}.
% It becomes possibly insecure because 
In Appendix~\ref{ssub:scheme 4} we have the condition $m> 2s$, thus the last two transmissions are guaranteed to not have common messages.
However, when the scheme is used in the case $m<2s$, it is possible that these two transmissions contain the common messages.
The users can possibly decode an extra message after decoding their desired messages thus making the scheme possibly insecure.

Specifically, when $m-s-2=2s-m+2a$ the last two transmissions have $w_{m-s-2}$ as the common message. 
User $u_{s+a_{1}-\floor{\frac{m}{2}}-1}$ has the desired message as $w_{m-s-2}$ and can decode $w_{m-s-2}$ by the last transmission.
Note that $a_{1}+\ceil{\frac{m}{2}}-1-( s+a_{1}-\floor{\frac{m}{2}}-1 )=m-s=p$. 
That means that user $u_{s+a_{1}-\floor{\frac{m}{2}}-1}$ can decode $w_{a_{1}+\ceil{\frac{m}{2}}-1}$ by the second transmission since there only two messages involved in the second transmission that are not in the side information set of user $u_{s+a_{1}-\floor{\frac{m}{2}}-1}$ and one of the messages is its desired message.
This violates the security constraint.

To secure the scheme, we choose $a$ such that $m-s-2\neq 2s-m+2a$, that is, $a\neq p-s/2-1$.
By such choice of $a$ the scheme becomes secure.
% Since $m-s-2\neq 2s-m+2a$ can not be true for 
If we have the freedom to choose $a$ to be either 2 or 3, the security constraint can be satisfied.
Therefore, the scheme does not work only if $a$ can only be $2$ and $p-s/2-1=2$.
That is, $7\leq {p}<9$ and $p=s/2+3$.
Two cases are included:
\begin{enumerate}
    \item $p=7, s=8, m=15$.
    \item $p=8, s=10, m=18$.
\end{enumerate}

\paragraph{Case $p=7, s=8, m=15$}
\label{para:separateCase_p=7,s=8,m=15}
We propose a scheme with $6$ transmissions that works for this case.
The transmissions are:
\begin{itemize}
    \item $\sum_{i=1}^{7}w_{i}$.
    \item $\sum_{i=3}^{9}w_{i}$.
    \item $\sum_{i=5}^{11}w_{i}$.
    \item $\sum_{i=7}^{13}w_{i}$.
    \item $w_{10}+w_{13}+w_{14}+w_{15}$.
    \item $w_{1}+w_{3}+w_{14}+w_{15}$.
\end{itemize}
The transmissions are illustrated in Fig.\ref{fig:scheme_p=7, s=8, m=15}.
\begin{figure}[ht]
    \centering
    \includegraphics[width=0.7\columnwidth]{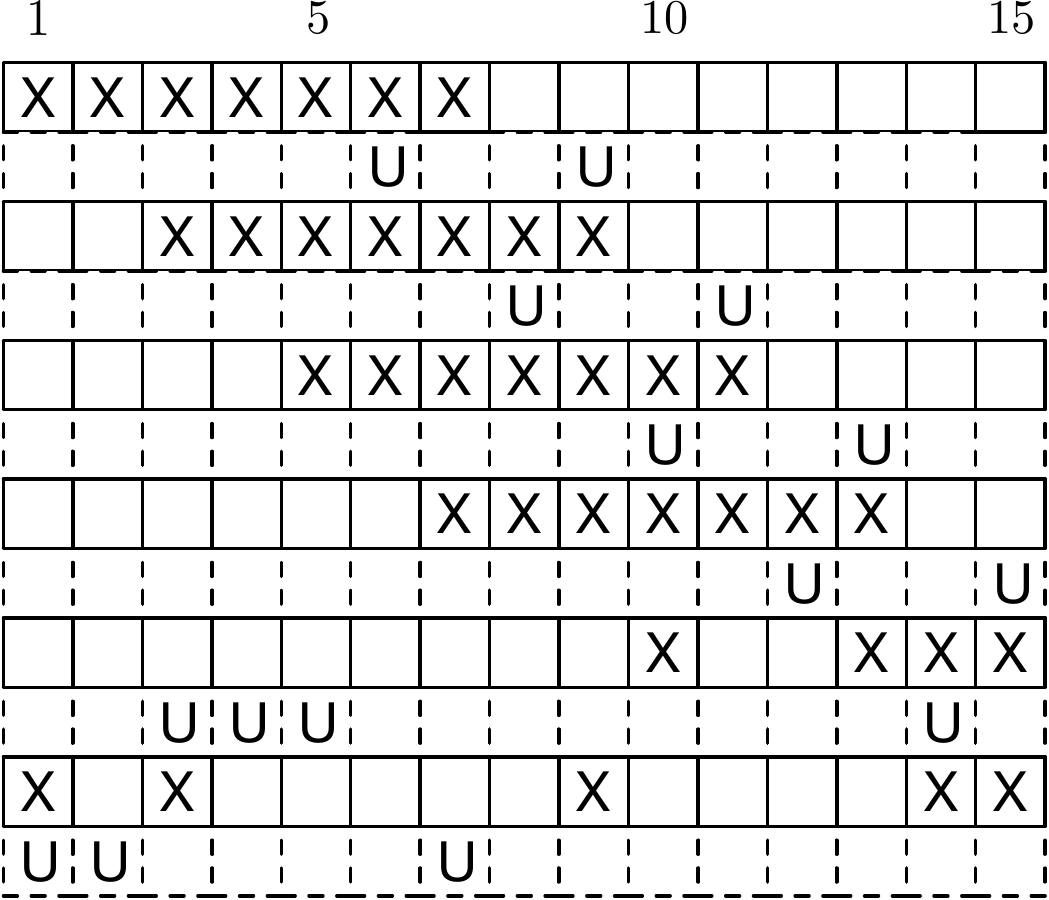}
    \caption{Scheme for the case $p=7, s=8, m=15$}
    \label{fig:scheme_p=7, s=8, m=15}
\end{figure}

\paragraph{Case $p=8, s=10, m=18$}
\label{para:separateCase_p=8,s=10,m=18}
Note that this is the case where $m$ and $s$ are even. 
By the result in~\cite{centralized_secure_picod} we know that there exists scheme with $m/2$ transmissions that can securely satisfy all users.
Since here we are only interested in a single case and we are satisfied by a scheme with a finite number of transmissions. 
The scheme with $m/2$ transmissions, although suboptimal, is enough for us to claim that for the case $p=8, s=10, m=18$ can be securely satisfied with a finite number of transmissions.

Therefore, for all cases with $p\geq 7$, we show the feasibility.

\subsection{Case $4\leq p\leq 6$}
\label{sub:case4<p<6}
We now study the remaining cases in the regime $3s/2<s<2s$, $ m-4 \geq s \geq 5$.

Note that in this regime we have $p<s<2p$.
Therefore, for $4\leq p\leq 6$, the number of possible cases are finite. 
We address them cases by case.

\paragraph{Even $m$}
\label{para:separateCase_even m}
The cases with even $m$ include $p=4,6$, even $s$ and $p=5$, odd $m$. 
The largest $m$ in this case is $p+s=6+10=16$.
By the result in~\cite{centralized_secure_picod}, the scheme with $m/2$ transmissions can securely satisfy these cases. 
The number of transmissions needed is upper bounded by $16/2=8$.

The remaining cases are the cases where $m$ is odd. 
\paragraph{Case $p=4, s=5, m=9$}
\label{para:separateCase_p=4,s=5,m=9}
We propose a scheme with $4$ transmissions that works for this case.
The transmissions are:
\begin{itemize}
    \item $\sum_{i=1}^{4}w_{i}$.
    \item $\sum_{i=3}^{6}w_{i}$.
    % \item $\sum_{i=5}^{11}w_{i}$.
    % \item $\sum_{i=7}^{13}w_{i}$.
    \item $w_{5}+w_{7}+w_{8}$.
    \item $w_{1}+w_{8}+w_{9}$.
\end{itemize}
The transmissions are illustrated in Fig.\ref{fig:scheme for p=4, s=5, m=9}.
\begin{figure}[ht]
    \centering
    \includegraphics[width=0.4\columnwidth]{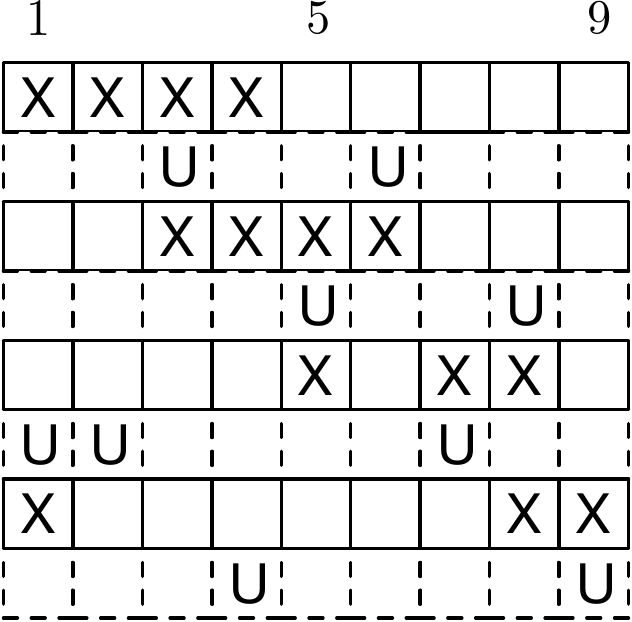}
    \caption{Scheme for the case $p=4, s=5, m=9$}
    \label{fig:scheme for p=4, s=5, m=9}
\end{figure}

\paragraph{Case $p=4, s=7, m=11$}
\label{para:separateCase_p=4,s=7,m=11}
We propose a scheme with $4$ transmissions that works for this case.
The transmissions are:
\begin{itemize}
    % \item $\sum_{i=1}^{6}w_{i}$.
    % \item $\sum_{i=3}^{8}w_{i}$.
    % \item $\sum_{i=5}^{10}w_{i}$.
    % \item $\sum_{i=7}^{13}w_{i}$.
    \item $w_{2}+w_{4}+w_{6}$.
    \item $w_{5}+w_{6}+w_{7}+w_{9}$.
    \item $w_{1}+w_{7}+w_{8}+w_{9}+w_{10}+w_{11}$
    \item $w_{1}+w_{2}+w_{3}+w_{10}+w_{11}$.
\end{itemize}
The transmissions are illustrated in Fig.\ref{fig:scheme for p=4, s=7, m=11}.
\begin{figure}[ht]
    \centering
    \includegraphics[width=0.5\columnwidth]{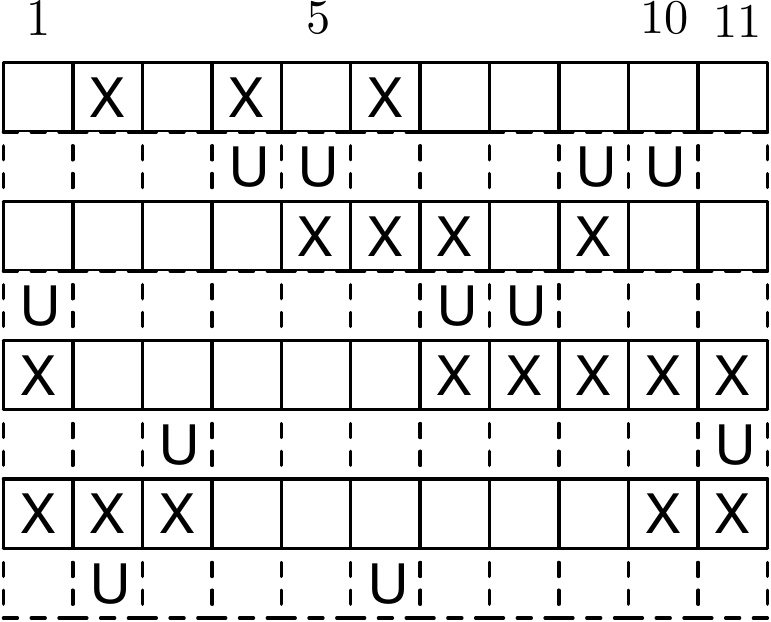}
    \caption{Scheme for the case $p=4, s=7, m=11$}
    \label{fig:scheme for p=4, s=7, m=11}
\end{figure}

\paragraph{Case $p=5,s=6,m=11$}
\label{para:separateCase_p=5,s=6,m=11}
We propose a scheme with $5$ transmissions that works for this case.
The transmissions are:
\begin{itemize}
    \item $\sum_{i=1}^{5}w_{i}$.
    \item $\sum_{i=3}^{7}w_{i}$.
    \item $\sum_{i=5}^{9}w_{i}$.
    % \item $\sum_{i=7}^{13}w_{i}$.
    \item $w_{7}+w_{9}+w_{10}+w_{11}$.
    \item $w_{1}+w_{2}+w_{11}$.
    % \item $w_{1}+w_{7}+w_{8}+w_{9}+w_{10}+w_{11}$
    % \item $w_{1}+w_{2}+w_{3}+w_{10}+w_{11}$.
\end{itemize}
The transmissions are illustrated in Fig.\ref{fig:scheme for p=5,s=6,m=11}.
\begin{figure}[ht]
    \centering
    \includegraphics[width=0.5\columnwidth]{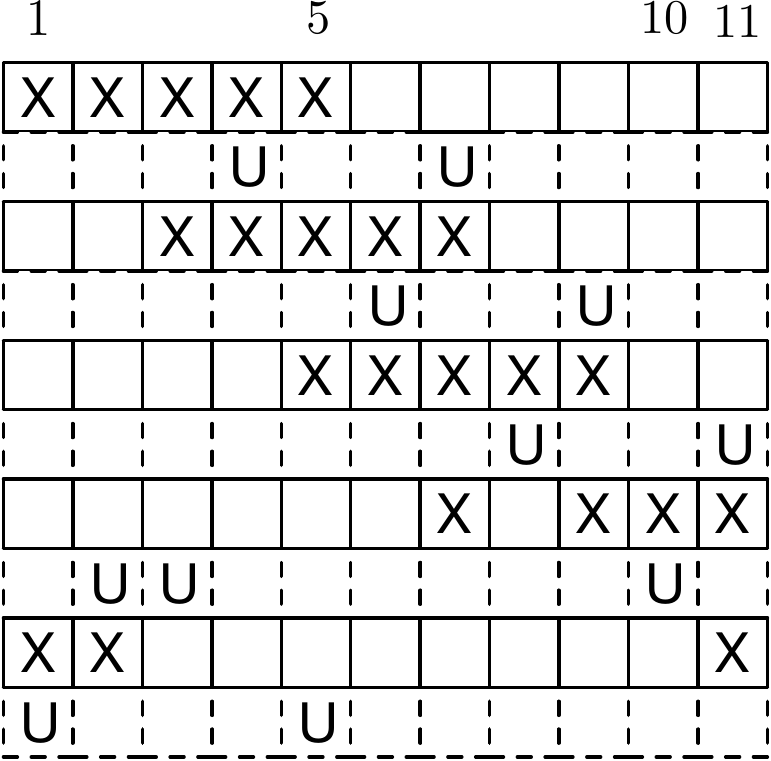}
    \caption{Scheme for the case $p=5,s=6,m=11$}
    \label{fig:scheme for p=5,s=6,m=11}
\end{figure}

\paragraph{Case $p=5,s=8,m=13$}
\label{para:separateCase_p=5,s=8,m=13}
We propose a scheme with $5$ transmissions that works for this case.
The transmissions are:
\begin{itemize}
    \item $\sum_{i=1}^{7}w_{i}$.
    \item $\sum_{i=3}^{9}w_{i}$.
    \item $\sum_{i=5}^{11}w_{i}$.
    % \item $\sum_{i=7}^{13}w_{i}$.
    % \item $w_{2}+w_{4}+w_{6}$.
    \item $w_{8}+w_{11}+w_{12}+w_{13}$.
    % \item $w_{1}+w_{7}+w_{8}+w_{9}+w_{10}+w_{11}$
    \item $w_{1}+w_{3}+w_{12}+w_{13}$.
\end{itemize}
The transmissions are illustrated in Fig.\ref{fig:scheme for p=5,s=8,m=13}.
\begin{figure}[ht]
    \centering
    \includegraphics[width=0.6\columnwidth]{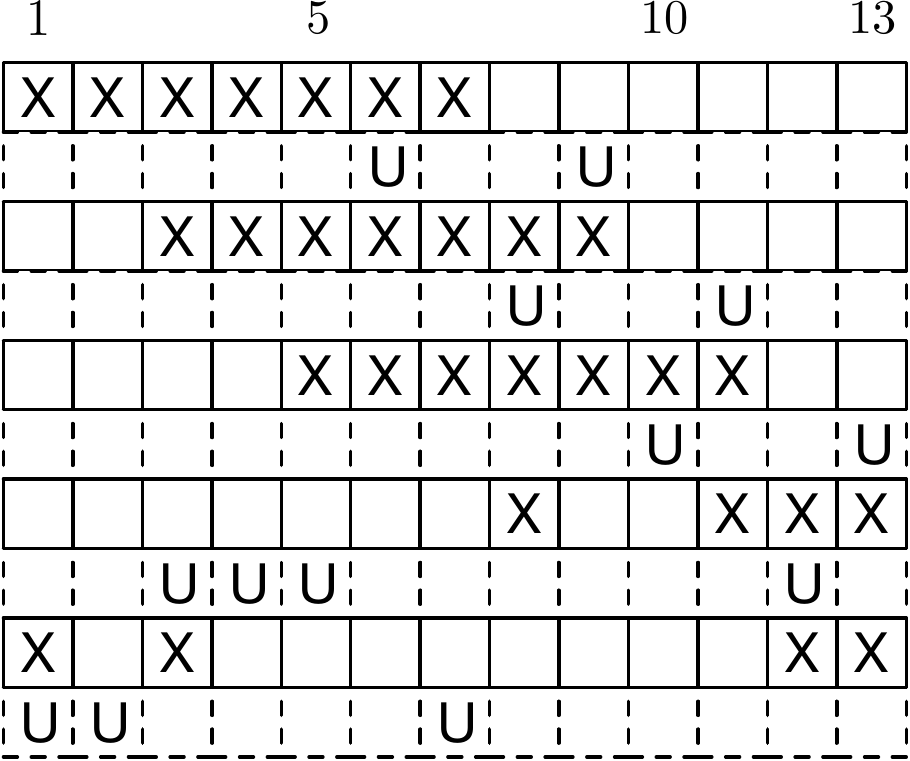}
    \caption{Scheme for the case $p=5,s=8,m=13$}
    \label{fig:scheme for p=5,s=8,m=13}
\end{figure}

\paragraph{Case $p=6,s=7,m=13$}
\label{para:separateCase_p=6,s=7,m=13}
We propose a scheme with $6$ transmissions that works for this case.
The transmissions are:
\begin{itemize}
    \item $\sum_{i=1}^{6}w_{i}$.
    \item $\sum_{i=3}^{8}w_{i}$.
    \item $\sum_{i=5}^{10}w_{i}$.
    \item $\sum_{i=7}^{12}w_{i}$.
    % \item $w_{2}+w_{4}+w_{6}$.
    % \item $w_{5}+w_{6}+w_{7}+w_{9}$.
    \item $w_{1}+w_{9}+w_{11}+w_{12}+w_{13}$
    \item $w_{1}+w_{2}+w_{3}+w_{12}+w_{13}$.
\end{itemize}
The transmissions are illustrated in Fig.\ref{fig:scheme for p=6,s=7,m=13}.
\begin{figure}[ht]
    \centering
    \includegraphics[width=0.6\columnwidth]{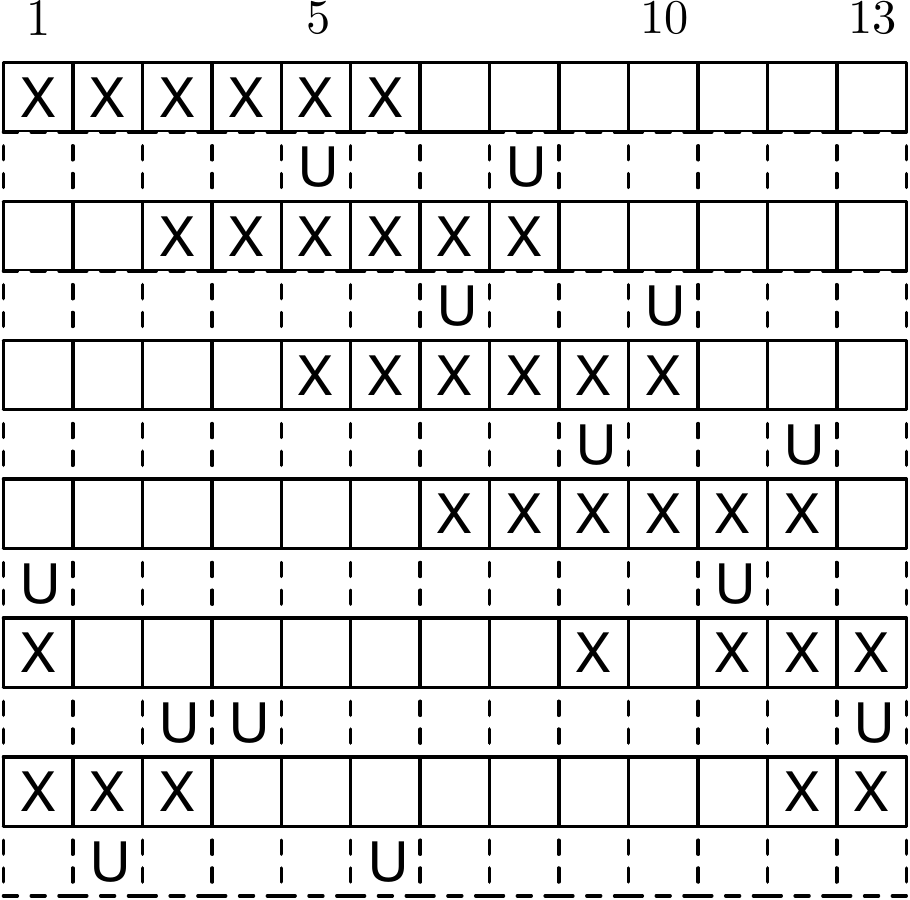}
    \caption{Scheme for the case $p=6,s=7,m=13$}
    \label{fig:scheme for p=6,s=7,m=13}
\end{figure}

\paragraph{Case $p=6,s=9,m=15$}
\label{para:separateCase_p=6,s=9,m=15}
We propose a scheme with $6$ transmissions that works for this case.
The transmissions are:
\begin{itemize}
    \item $\sum_{i=1}^{8}w_{i}$.
    \item $\sum_{i=3}^{10}w_{i}$.
    \item $\sum_{i=5}^{12}w_{i}$.
    \item $\sum_{i=7}^{14}w_{i}$.
    % \item $w_{2}+w_{4}+w_{6}$.
    % \item $w_{5}+w_{6}+w_{7}+w_{9}$.
    \item $w_{1}+w_{10}+w_{13}+w_{14}+w_{15}$
    \item $w_{1}+w_{2}+w_{4}+w_{14}+w_{15}$.
\end{itemize}
The transmissions are illustrated in Fig.\ref{fig:scheme for p=6,s=9,m=15}.
\begin{figure}[ht]
    \centering
    \includegraphics[width=0.7\columnwidth]{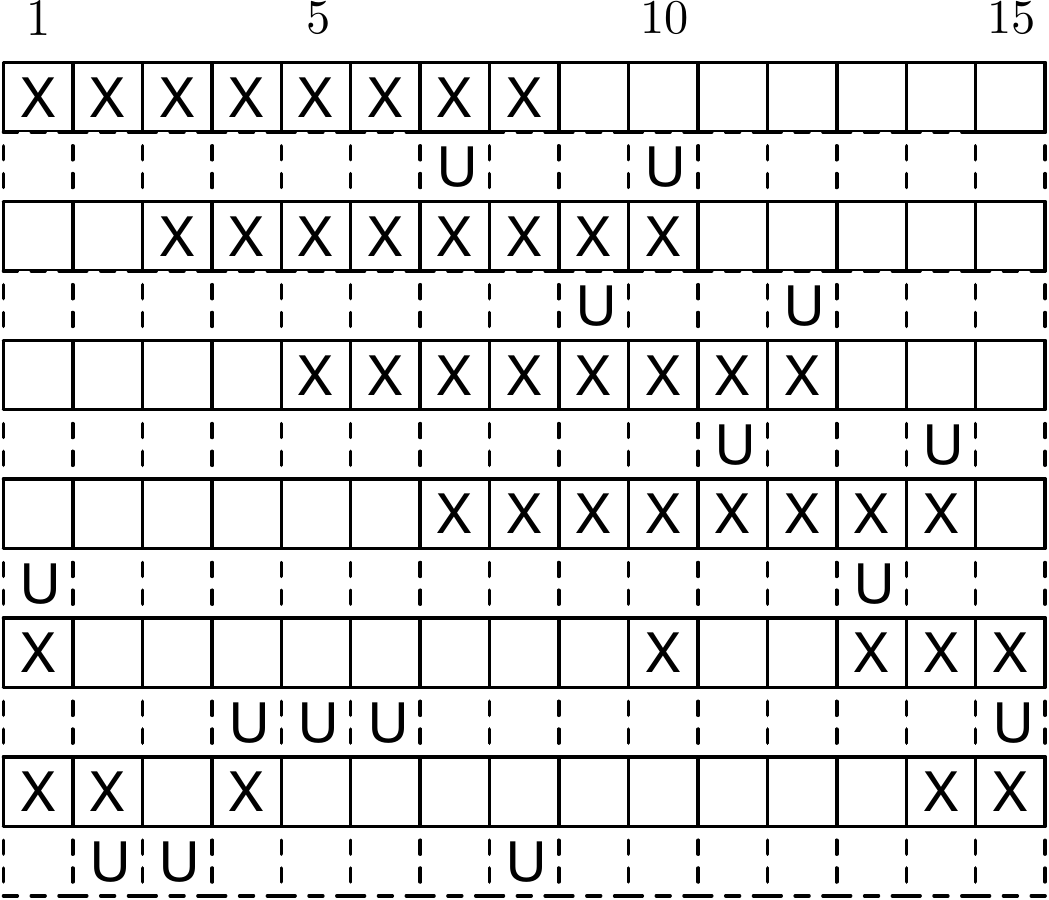}
    \caption{Scheme for the case $p=6,s=9,m=15$}
    \label{fig:scheme for p=6,s=9,m=15}
\end{figure}

\paragraph{Case $p=6,s=11,m=17$}
\label{para:separateCase_p=6,s=11,m=17}
We propose a scheme with $6$ transmissions that works for this case.
The transmissions are:
\begin{itemize}
    \item $\sum_{i=1}^{10}w_{i}$.
    \item $\sum_{i=3}^{12}w_{i}$.
    \item $\sum_{i=5}^{14}w_{i}$.
    \item $\sum_{i=7}^{16}w_{i}$.
    % \item $w_{2}+w_{4}+w_{6}$.
    \item $w_{1}+w_{1}+w_{15}+w_{16}+w_{17}$.
    % \item $w_{1}+w_{7}+w_{8}+w_{9}+w_{10}+w_{11}$
    \item $w_{1}+w_{2}+w_{5}+w_{16}+w_{17}$.
\end{itemize}
The transmissions are illustrated in Fig.\ref{fig:scheme for p=6,s=11,m=17}.
\begin{figure}[ht]
    \centering
    \includegraphics[width=0.8\columnwidth]{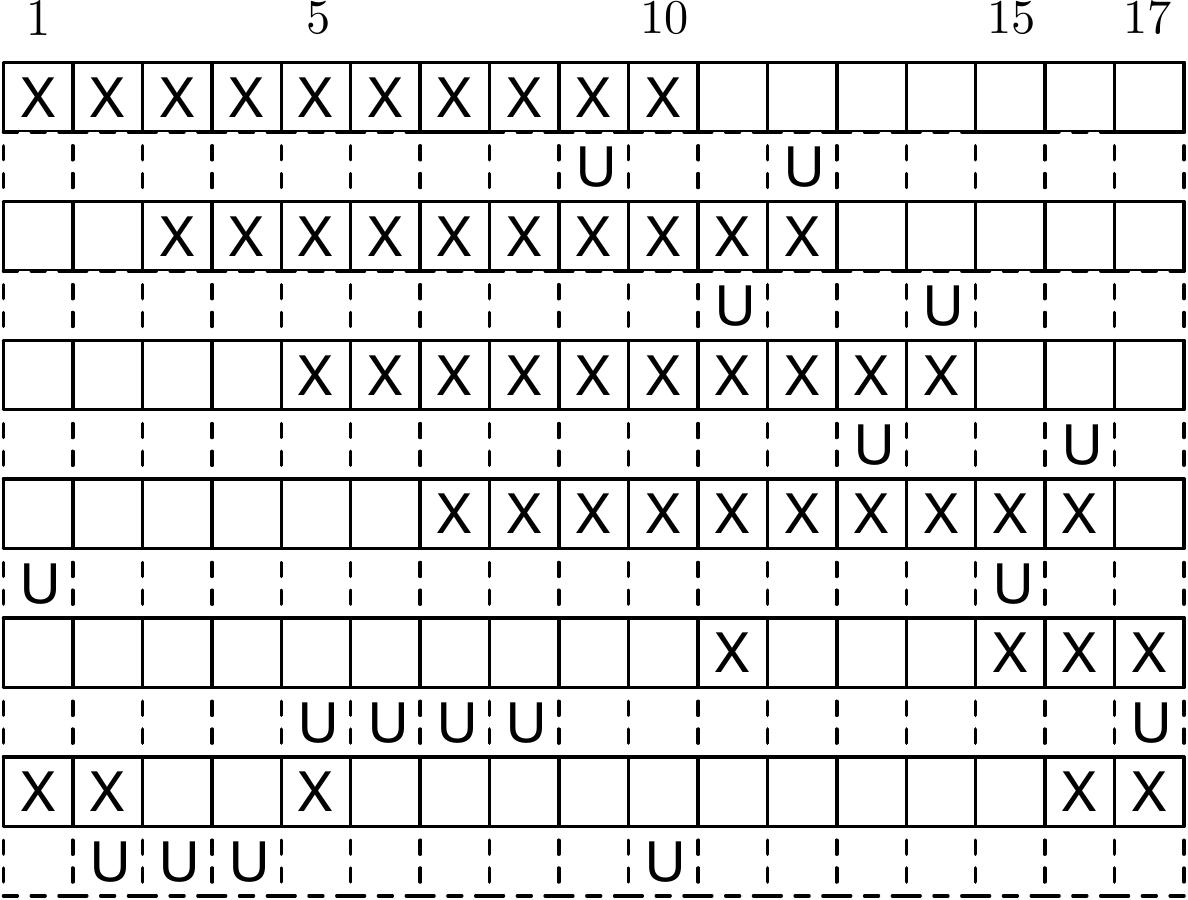}
    \caption{Scheme for the case $p=6,s=11,m=17$}
    \label{fig:scheme for p=6,s=11,m=17}
\end{figure}

We address all cases in the regime where  $3s/2<m < 2s$, $ m-4 \geq s \geq 5$, and $4\leq m-s \leq 6$ and show all cases are feasible.
In this regime, the maximum number of transmissions needed is $9$ for the case $m=18, s=10$.

\section{Case $s\leq 4$ or $m-s=p\leq 3$}
\label{sec:small large s}

In this section we consider the cases where the size of the side information set $s$ is ``small'' or ``large'' when compared to the number of users $m$.
These cases are studied separately because this regime contains all infeasible cases. 
Yet not all cases in this regime are infeasible. 
Specifically, the feasible cases include $s=3,4$, even $m$ and $s=m-3$.

\subsection{$s=3$ and even $m$}
\label{sub:s=3 and even m}

In this case we use the scheme that takes $m/2$ transmissions in~\cite{centralized_secure_picod}.
This scheme is actually optimal in this case since $3m/2s=m/2$ in this case. 

\subsection{$s=4$ and even $m$}
\label{sub:s=4 and even m}

Note that $m \bmod{2s} \in \{0,2,4,6\}$ in this case. 
Therefore, by applying scheme 1 in Appendix~\ref{ssub:scheme 1}, there will be $0$ or $2$ or $4$ or $6$ users remain. 
Then by scheme 4 in Appendix~\ref{ssub:scheme 4} we can satisfy the remain users with $0,1,2,3$ transmissions, respectively. 
Note that in this case the condition $s\geq 5$ on scheme 4 can be relaxed  since the number of remaining users is always even so that the number of users need to be satisfied by scheme 4 will never be $s+1$. 
% The requirement $s\geq 5$ is for the scheme to satisfy $s+1$ users. 
% Therefore the requirement can be relaxed to $s\geq 4$, which is the requirement for the scheme to satisfy $s+2=6$ users.
Also, note that in this case $s+1>2s-4$, therefore the last subcase in scheme 4 will never happen.

% \subsubsection{$s<5$}
% So far we have the achievable schemes for all cases where $s\geq 5$. 
% We now deal with the remaining cases where $s<5$.
% In this case the side information set is small.  
% Many cases are infeasible. 
% Section~\ref{sec:infeasibility} shows that without 1-factor, the cases $s=1,2$ are infeasible. The cases $s=3,4$ are infeasible when $m$ is odd. 
% We now check the feasible cases, that is, $s=3,4$ and $m$ is even.

% For the case $s=3$ and even $m$, we use scheme 6 as the achievable scheme. 
% The scheme takes $m/2+2-\ceil{3/2}=m/2$ transmissions. It coincides with the linear converse bound, therefore is linearly optimal

\subsection{$s=m-3$ and even $m$}
\label{sub:s=m-3 and even m}

3 transmissions can securely satisfy all users: $\sum_{i=1}^{s-1} w_i, \sum_{i=3}^{s+1} w_i, \sum_{i=5}^{m} w_i$.
Note that by this scheme every user decodes one message, which is its desired message, and a sum of the other two messages that are not in its side information.

\subsection{$s=m-3$ and odd $m$}
\label{sub:s=m-3 and odd m}

% \subsubsection{}
% {\color{red}
% This scheme is obsolete.}
Note that $m=5,7$ are infeasible as shown in Section~\ref{sec:infeasibility}.
$m=9$ is the case $\frac{m}{m-s}=\frac{9}{3}=3\in\mathbb{Z}$.
Therefore, this case contains odd $m\geq 11$.
We use the scheme for odd $m=s+3$ and $m\geq 11$ proposed in\cite{centralized_secure_picod}.
We satisfy all users by 4 transmissions. The 4 transmissions are  $\{ w_1+\sum_{i=1}^{\frac{m-7}{2}}w_{2i}+w_{m-6}, \sum_{i=1}^{\frac{m-9}{2}} w_{2i+1}+w_{m-5}+w_{m_4}, w_{m-3}+w_{m-2}, w_{m-1}+w_{m} \}$.

% \section{Large $s$}
% \label{sec:large s}

\end{document}